\documentclass[prx,10pt,notitlepage,twocolumn,superscriptaddress]{revtex4}

\usepackage{graphicx}
\usepackage{amsmath}
\usepackage{amssymb}
\usepackage{amsfonts}
\usepackage{color}
\usepackage{xspace}
\usepackage{subfigure}
\usepackage{bm}
\usepackage{bbold}
\usepackage{blkarray}
\usepackage{pdflscape}
\usepackage{widetable}
\usepackage{enumerate}

\usepackage{blkarray}

\newcommand{\tJ}{$t$-$J$\xspace}
\newcommand{\bra}[1]{\left\langle #1 \right|}
\newcommand{\ket}[1]{\left| #1 \right\rangle}
\newcommand{\bral}[1]{\langle #1 |}
\newcommand{\ketl}[1]{| #1 \rangle}

\newcommand{\Mo}{Mo$_3$S$_7$(dmit)$_3$\xspace}
\newcommand{\etal}{\emph{et al}. }

\begin{document}

\title{Spin-orbit coupling and strong electronic correlations in cyclic molecules}

\author{A. L. Khosla}
\email{amie.khosla@uqconnect.edu.au}
\affiliation{School of Mathematics and Physics, The University of Queensland,  Queensland, 4072, Australia}

\author{A. C. Jacko}\affiliation{School of Mathematics and Physics, The University of Queensland,  Queensland, 4072, Australia}

 \author{J. Merino}
 \affiliation{Departamento de F\'isica Te\'orica de la Materia Condensada, Condensed Matter Physics Center (IFIMAC) and
 	Instituto Nicol\'as Cabrera, Universidad Aut\'onoma de Madrid, Madrid 28049, Spain}

\author{B. J. Powell}\affiliation{School of Mathematics and Physics, The University of Queensland,  Queensland, 4072, Australia}

\begin{abstract}
In atoms spin-orbit coupling (SOC) cannot raise the angular momentum above a maximum value or lower it below a minimum. 
Here we show that this need not be the case in materials built from  nanoscale structures including multi-nuclear coordination complexes, materials with decorated lattices, or atoms on surfaces. In such cyclic molecules the electronic spin couples to currents running around the molecule. For  odd-fold symmetric molecules (e.g., odd membered rings) the SOC is highly analogous to the atomic case; but for even-fold symmetric molecules every angular momentum state can be both raised and lowered. These differences arise because for odd-fold symmetric molecules the maximum and minimum molecular orbital angular momentum states are time reversal conjugates, whereas for even-fold symmetric molecules they are aliases of the same single state.  We show, from first principles calculations, that in suitable molecules this molecular SOC is large, compared to the energy differences between frontier molecular orbitals. Finally, we show that, when electronic correlations are strong, molecular SOC can cause highly anisotropic exchange interactions and discuss how this can lead to effective spin models with compass Hamiltonians.
	%
\end{abstract}
\maketitle

\section{Introduction}

Electrons traveling at relativistic velocities  experience a spin-orbit coupling (SOC): $H_\textrm{SO}=\bm K\cdot\bm \sigma$, where $\bm \sigma$ is the spin operator. The properties of the pseudovector $\bm K$ depend on the symmetry of the system. 
For spherical symmetry, e.g., in atoms, $\bm K=\lambda\bm L$, where $\lambda$ is a constant and $\bm L$ is the orbital angular momentum. But in   lower symmetry environments  SOC can be rather different. Important instances of this were discovered by Dresselhaus and Rashba \cite{Manchon,Bihlmayer}. 

In spherically symmetric systems there is a maximum (minimum) state that cannot be surpassed by applying a raising (lowering) angular momentum operator.
This  constrains  which states are coupled by SOC \cite{Sakurai}. 
{However, we will see below, that in systems built from nanoscale structures  these constraints are modified and very different spin-orbit Hamiltonians are realized. We will consider systems where the internal energy scales within the nanostructure are large compared to the intra-structure energy scales, such that one may integrate many internal degrees of freedom out of a low-energy effective Hamitlonian.}

In molecular crystals the building blocks are fundamentally the molecules themselves \cite{KK,JPCM}. We will argue below that  multi-nuclear coordination complexes \cite{LlusarReview,LlusarJACS,Almeida} provide an ideal platform to explore our ideas. But, we stress that our results apply equally to any other system with the same symmetry, for example, arrays of heavy atoms arranged into polygons on a surface {\cite{Loth,Fiete,Crommie} or  materials that form decorated lattice models \cite{Bao,Zhong,Oakley,Sheckelton12,Sheckelton14,Mourigal}. Materials built from such nanostructures contain a hierachy of energy scales that  makes them particularly flexible platforms for  engineering  specific SOC Hamiltonians tuned to different applications. }

{In this paper, we focus on the most physically transparent version of the problem:   molecules with $N$-fold cyclic symmetry ($\mathcal{C}_N$). Because the spherical symmetry of the atoms is strongly broken in the nanostructured environment it is not correct to view the SOC as a linear combination of the SOC on the atoms that make up the molecule. Rather, there is an emergent `spin molecular orbital coupling' (SMOC)  that couples the electronic spin  to currents running around the molecule. This result is quite general and the methodology described below can be  extended to molecules or nanostructures with  symmetries other than those discussed here.
}

For odd $N$-fold symmetric molecules (e.g., odd membered rings)  the consequences of SMOC is highly analogous to the atomic SOC; but for even membered rings every angular momentum state can be both raised and lowered. We present density  functional calculations that identify specific multi-nuclear organometallic complexes where the SMOC is large compared to other relevant energy scales. {We show that our postulated form of the SMOC arises in the $\mathcal{C}_3$ symmetric molecule \Mo  from these unbiased \textit{ab initio} calculations.} Finally, we explore a potential application of our findings: controlling the anisotropy of magnetic exchange interactions in systems where electronic correlations are strong. We show that the interplay of  SMOC with electronic correlations can give rise to effective spin models with compass Hamiltonians. These models are known to give rise to many interesting states of matter, including some with topological order. Unlike previous schemes to realize such Hamiltonians, these effects do not rely on hopping through intermediate atoms or molecules \cite{Jackeli,Oshikawa,Perkins,Bendix}. This provides an example of how the SOC can be controlled in molecular materials and engineered for a specific application. 

Potential applications of designer SOC include molecular qubits, spintronics and organic electronics \cite{Shiddiq,Freedman,Manchon,Bihlmayer,CCR}.
Furthermore, strong SOC is required to realize many symmetry protected topological phases of matter, such as topological insulators and superconductors, quantum spin Hall states, axion insulators and Weyl semimetals \cite{Qi,Xu,Wan,BalentsARCMP,Soluyanov}.



When both SOC and electronic correlations are strong  additional phases are possible, including topological Mott and Kondo insulators \cite{Balents,Galitski}.
Moreover, these ingredients allow for true topological order, which is characterized by long range-entanglement and often supports fractionalized quasiparticles \cite{Wen,Castelnovo}. Interest in this physics was redoubled by Kitaev's exact solution \cite{Kitaev} of  a compass model, \textit{i.e.}, a spin model with exchange interactions that are highly anisotropic in both real- and spin-space \cite{Brink}. Kitaev found a topological spin liquid with
non-abelian anyonic excitations, which is sufficient to enable fault tolerant quantum computation \cite{Pachos}. Jackeli and Khaliullin \cite{Jackeli} argued that the low-energy physics of a class of iridiumoxides (iridates) are described by the Kitaev model because of their strong SOC.  However, it was soon realized that in this picture there must be a large isotropic exchange interaction \cite{Kee}. Indeed, it has been  argued that  the Kitaev model does not describe the iridates \cite{Mazin}. This has  renewed the search for materials that may realize the physics of the Kitaev and other compass models \cite{Bendix,Oshikawa}. However, in previous proposals the SOC arises from intra-atomic SOC on a transition metal, which  is surrounded by multiple light atoms, thus the SOC is essentially atomic \cite{A2B2}. 


In molecular crystals the electrons hop between molecular orbitals which are significantly larger than the atomic orbitals relevant in, say, transition metal oxides. This leads to an effective on-site Coulomb interaction, $U$, that is typically an order of magnitude smaller than  in transition metal oxides  \cite{JPCM,KK}. However, the intra-molecular hopping integral, $t$, is also typically an order of magnitude smaller. This means that electrons in molecular crystals are typically strongly correlated. Furthermore, this implies that {the strength of the SOC can be large relative to other relevant energy scales} in  molecular crystals.

{In atoms the SOC increases with the atomic number. This remains true in molecules as heavier atoms imply larger gradients in the nuclear potential, cf. Eq. (\ref{Pauli}).}
In   organic materials the largest contributions to  SOC typically arise from sulfur  or selenium atoms \cite{WinterReview}. Therefore, a powerful strategy for increasing SOC is  to move to organometallic complexes; this has driven much recent progress in organic solar cells and organic light-emitting diodes  \cite{CCR}. Therefore, multi-nuclear organometallic complexes (i.e., molecules containing multiple transition metal atoms) with ligands that facilitate effective charge transport between molecules  \cite{LlusarReview,LlusarJACS,Almeida} provide a platform  that allows for synthetic control and engineering  of SMOC beyond the possibilities available in inorganic systems. Furthermore, these materials will facilitate new ways to explore the interplay of the SMOC with strong electronic correlations. As an example of this, we propose  that compass models can be realized in crystals of multi-nuclear organometallic complexes. 

\section{Spin-orbit Hamiltonian in cyclic molecules}

A variety of low-velocity approximations to the Dirac equation can be constructed, such as the Pauli, Briet-Pauli and regular approximations \cite{Dyall}. The details of the pseudovector $\bm K$ in the SOC Hamiltonian, $H_{SO}={\bm K}\cdot{\bm \sigma}$, depend to some extent on which approximation is chosen. However, in what follows we will only make use of the symmetries of $\bm K$, which are independent of the low-velocity approximation as they are inherited from the Dirac equation.
In many low-velocity approximations one can write
\begin{eqnarray}
\bm K&=&\frac{\hbar}{4m^2c^2}(\bm p\times\bm\nabla V(\bm r)),
\label{Pauli}
\end{eqnarray}
where  $\bm p$  is the momentum  operator, and  $V(\bm r)$ is a (screened) potential \cite{Dyall}.
{Thus, in  a molecule $V(\bm r)$ is simply a linear superposition of the atomic potentials. But in molecular systems one expects that the expectation values of $\bm p$ will be very different from those for electrons orbiting a single atom, particularly for states near the Fermi energy. Thus, in molecules it is not, in general, correct to assume that the SOC is simply a  linear superposition of the atomic SOC ($\lambda{\bm L}\cdot{\bm S})$ \cite{Ref1,Ref2,Ref3}. Indeed, we will show below that this assumption would lead to the neglect of important physics.}

The standard approaches to this problem in molecular systems are either to evaluate {the matrix elements of the full  $\bm K$ operator} from first principles \cite{WinterReview,CCR,Ref6,Hess,Neese} or to assume that only the SOC on selected heavy atoms is relevant and the SOC retains the spherical symmetry of the atomic case {on those heavy atoms} \cite{CCR,Kee,Jackeli,MacDonald,Baartas}. {The former approach has been widely applied to both organic and organometallic molecules while the latter approach has found wide applications in materials systems such as transition metals oxides and mononuclear coordination complexes.} Here we take the alternative approach of simply analyzing which matrix elements are allowed in arbitrary molecules with cyclic,  $\mathcal{C}_N$, symmetries.
This analysis will make extensive use of the cyclic double groups $\tilde{\mathcal{C}}_N$,  Table \ref{tab:odd}.

\begin{table}[t]
	\centering
	\begin{tabular}{l|cccc|l} 
		{\bf Odd} ${\bm N}$ & $E$ & $(C_N)^n$ &  $\bar{E}$ & $(\bar{C}_N)^n$ & TR \\ \hline 
		$A_0$         & 1   & 1         &  1 &    1   & ${\cal T}|0\rangle=|0\rangle$ \\ 
		$E_{k}$       & 1   & $\omega^{kn}$    &  1 & $\omega^{kn}$ & ${\cal T}|k\rangle=(-1)^k|-k\rangle$\\
		\hline 
		$E_{j}$     & 1   &  $\omega^{jn}$   & $-1$ & $-\omega^{jn}$     & ${\cal T}|j\rangle=(-1)^{j-\frac12}|-j\rangle$\\
		$A_{N/2}$     & 1  &   $(-1)^n$      & $-1$ &  $(-1)^{n-1}$     &   ${\cal T}|N/2\rangle=|N/2\rangle$
		\\\hline\hline
		\textbf{Even} ${\bm N}$ & $E$ & $(C_N)^n$ &  $\bar{E}$ &  $(\bar{C}_N)^{n}$ & TR 
		\\ \hline 
		$A_0$         & 1   & 1         &  1 &       1        & ${\cal T}|0\rangle=|0\rangle$ \\ 
		$E_{k}$       & 1   & $\omega^{kn}$    &  1 & $\omega^{kn}$  & ${\cal T}|k\rangle=(-1)^k|-k\rangle$ \\
		$B_{N/2}$     & 1   & $-1$         &  1 &     $-1$        & ${\cal T}|N/2\rangle=|N/2\rangle$ \\  \hline
		$E_{j}$     & 1   &  $\omega^{jn}$   & $-1$ &   $-\omega^{jn}$   &${\cal T}|j\rangle=(-1)^{j-\frac12}|-j\rangle$\\
	\end{tabular}
	\caption{Character tables \cite{Koster,D2} for  the double groups $\tilde{\mathcal{C}}_N$.
		For a given $N$, representations `above the line' describe bosonic states (including even numbers of fermions), while those below the line are fermionic representations.
		The names of the representations, $A$, $B$ and $E$, are chosen in accordance with   Schoenflies notation. The additional subscript denotes angular momentum about the $C_N$ axis associated with the states that transform according to the representation.
		The operations of the single group are the identity, $E$, and rotation by $2\pi n/N$, $(C_N)^n$. The additional operations of the double group are indicated by a bar above these operations, implying a further rotation by $2\pi$.  
		Group multiplication simply adds the subscripts with periodic boundary conditions such that the sum lies in the interval $(-N/2,N/2]$. 
		The rightmost column indicates the behavior of a typical state that transforms according to the given representation under time reversal.
		For $N\geq3$ $S^z$, $S^+$, and $S^-$ are bases of $A_0$, $E_1$, and $E_{-1}$ respectively. 
		Here $1\leq n\leq N-1$ and $\omega=\exp(i 2\pi/N)$.
		For odd $N$,  $(1-N)/2\leq k\leq(N-1)/2$ and $-\frac{N}{2}< j\leq\frac{N}{2}$.
		For  even $N$, $-\frac{N}{2}< k\leq\frac{N}{2}$ and $(1-N)/2\leq j\leq(N-1)/2$.  $k$ is integral and $j$ is half-odd-integral for all $N$. $k=0$ refers to the representation $A_0$ and  $j=N/2$ ($k=N/2$) refers to $A_{N/2}$ ($B_{N/2}$).
	}
	\label{tab:odd}
\end{table}

It is convenient to introduce an orthogonal set of single electron basis states. The $\mu$th basis state in the fermionic representation $\overline\Gamma_j$   can be written as $|\overline{j}_{\mu}\rangle=|\underline{k}_\nu;\sigma\rangle\equiv|\underline{k}_\nu\rangle\otimes|\sigma\rangle$,  where the molecular orbital part of the wavefunction,   $|\underline{k}_\nu\rangle$, is the $\nu$th basis state that transforms as $\underline\Gamma_k$, a bosonic representation with integer $k$, and the spin part, $\ket{\sigma}$, transforms as $\ketl{\uparrow}\in E_{1/2}$ and $\ketl{\downarrow}\in E_{-1/2}$ for the non-trivial cyclic groups.

SMOC obeys a set of selection rules, which are derived in Appendix \ref{sect:rules}: 
\begin{subequations}
(1) SMOC does not couple time reversed states: 
\begin{eqnarray}
\bra{\overline{j}_\mu}H_{\mathrm{SO}}{\cal T}\ket{\overline{j}_\mu} =0, \label{eq:Krammer}
\end{eqnarray}
where ${\cal T}$ is the time reversal operator. 
This is a corollary to  Kramers' theorem  \cite{D2}.
(2) States with the same spin are only coupled by SMOC if their orbital parts belong to the same irreducible representation:
\begin{eqnarray}
\Big\langle{\underline{k}_\mu;\sigma}\Big|H_{\mathrm{SO}}\ket{\underline{q}_\nu;\sigma} 
=\sigma\lambda_{k;\mu\nu}^z\delta_{kq},
\label{eq:lz}
\end{eqnarray}
where $\lambda_{k;\mu\nu}^z=\lambda_{k;\nu\mu}^{z*}=-\lambda_{-k;\mu\nu}^z$ is a constant and $\sigma=\pm1/2$. (3) States with opposite spins are only coupled by SMOC if this conserves 
 $j=k+\sigma$:
\begin{eqnarray}
	\bra{\underline{q}_\nu;-\sigma}H_{\mathrm{SO}}\Big|{\underline{k}_\mu;\sigma}\Big\rangle = 
\frac12\lambda_{k+\sigma;\mu\nu}^{\pm}\delta_{k,q-2\sigma},
\label{lxy}
\end{eqnarray}
 where $\lambda_{k+\frac12;\mu\nu}^{\pm}$ is a  constant.
\label{sel_rule}
\end{subequations} 
Note that
these selection rules are quite natural if one interprets $k$  as the molecular angular momentum about the  $C_N$ axis,  henceforth  the $z$-axis, and $j$ as the total angular momentum about $z$.

As  Eqs. (\ref{sel_rule}) only depend on the symmetries of the  Hamiltonian  multiple low-velocity approximations to the Dirac equation \cite{Dyall}  yield the same selection rules. For example, Eqs. (\ref{sel_rule}) can be derived from, e.g., the Briet-Pauli formalism if  the two-electron SOC is treated at the mean-field level  \cite{Hess,Neese}.

These selection rules have surprising consequences in cyclic molecules.
To illustrate this, we consider the simplest class of models, where the low-energy physics is described by  $N$ orbitals  related by the cyclic symmetry described by the group ${\mathcal{C}}_N$, e.g., the one-band tight-binding, Hubbard, and \tJ models. The assumption that only a single orbital is relevant to each heavy atom is natural for the transition metals in multi-nuclear organometallic complexes. Typically, in such molecules the transition metals sit in low-symmetry environments, thus often the degeneracy of the atomic $d$-orbitals will be completely lifted.

The $\tilde{\mathcal{C}}_N$ tight-binding model is diagonalized by a Bloch transformation.
However, ${\cal T}^{-1}S^\pm{\cal T}=-S^\mp$ and $H_\textrm{SO}$ is time-reversal symmetric; implying that ${\cal T}^{-1}K^\pm{\cal T}=-K^\mp$, where $K^\pm\equiv K^x\pm iK^y$. To avoid  phase factors in the operators it is convenient to absorb them into the basis states:
\begin{eqnarray}
\ket{k}=\eta_k\sum_r e^{i\phi kr}\ket{r},
\label{Eq:Fourier}
\end{eqnarray}
where $-N/2<k\leq N/2$ and  $0\leq r\leq N-1$ are integers,  $\ket{r}$ is a Wannier orbital centered at $r$,  $\phi=2\pi/N$ and $\eta_k$ is a phase factor.
For  SO(3) symmetry the $\eta_k$ are usually chosen following the Condon-Shortley  convention, $\eta_k=i^{|k|}i^k$, cf. {\it e.g.}, the spherical harmonics. However, this does not respect time reversal symmetry for `aromatic' systems, where $N=4n+2$ for integer $n$. Therefore, we set $\eta_k=i^{|k|}$, which introduces the required phases for arbitrary $N$.
The state $\ket{k}$ is a basis for $\underline\Gamma_k$ and describes a (spinless)  current running around the molecule with angular momentum $\hbar k$. 
 
Applying the selection rules  [Eqs. (\ref{sel_rule})],  one finds that for  odd $N$
\begin{eqnarray}
H_{\mathrm{SO}}
&=& 
\sum_{m=1}^{L}\sum_{\sigma=-1/2}^{1/2}\sigma\lambda^z_m\left(\hat c_{m\sigma}^\dagger \hat c_{m\sigma} - \hat c_{-m\sigma}^\dagger \hat c_{-m\sigma} \right)
\label{eq:Hodd}
\\&&
\notag\hspace*{-0.2cm}+
\frac{1}{2} \sum_{j=\frac12}^{L-\frac12}\left[\lambda_j^{\pm}\left(\hat c_{j+\frac12\downarrow}^\dagger \hat c_{j-\frac12\uparrow} 
+\hat c_{-j+\frac12\downarrow}^\dagger \hat c_{-j-\frac12\uparrow} \right)
+\textrm{H.c.}\right]
\end{eqnarray}
where $\lambda^z_m$ is real and $\lambda^z_0=0$  by Eq. (\ref{eq:lz}), $N=2L+1$ implying $L\in\mathbb{Z}$,  $\hat c_{k\sigma}^\dagger$ creates an electron in the state $\ket{k;\sigma}$, which transforms according to the representation $\overline\Gamma_{k+\sigma}$, and sums in subscripts are defined modularly on the half-odd-integers  $(-N/2,N/2]$. 

Kramers' theorem [via Eq. (\ref{eq:Krammer})]  implies that  matrix elements between time reversed fermionic states  vanish -- importantly for odd $N$ this includes 
$\bra{-L;\downarrow}H_{\mathrm{SO}}\ket{L;\uparrow}$
even though both $\ket{-L;\downarrow}$ and $\ket{L;\uparrow}$ transform according to  $A_{N/2}$. 
Thus we find that, up to the values of matrix elements, which are not  determined by  symmetry, in the odd-$N$ case the structure of Eq. (\ref{eq:Hodd}) is equivalent to that in an atomic orbital with angular momentum $L$, where $H_\textrm{SO}^\textrm{at}=\lambda\bm L\cdot\bm S$. However, in general, the values of the constants ($\lambda_m^z$ and $\lambda_j^\pm$) break the spherical symmetry.

\begin{figure}
	\begin{center}
		\includegraphics[width=\columnwidth]{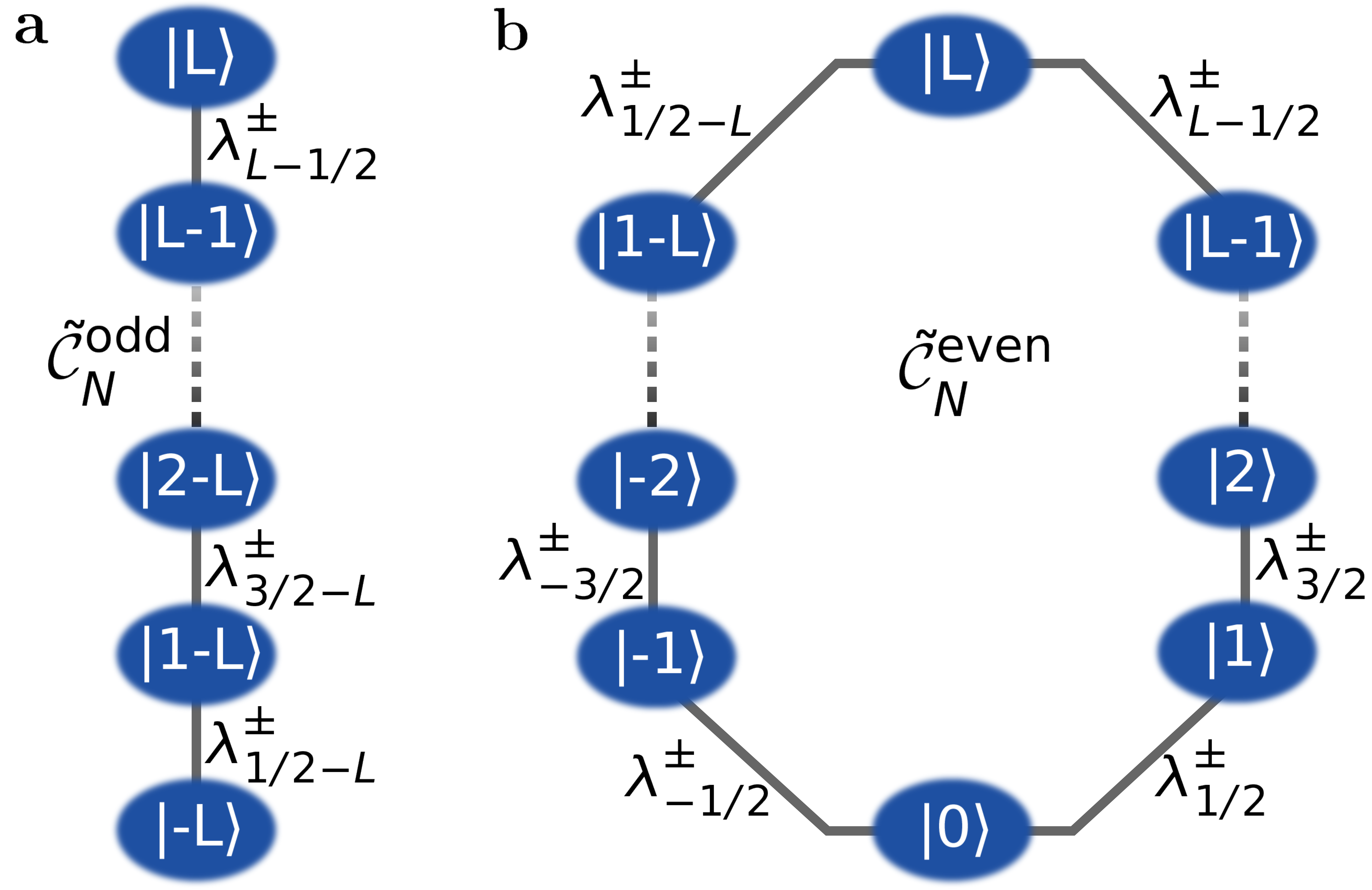}
	\end{center}
	\caption{Allowed matrix elements of $H_{SO}$ for systems with  cyclic symmetry, $\tilde{\cal C}_N$. (a) For odd $N$ there is a maximum (minimum) molecular angular momentum state $\ket{L}$ ($\ket{-L}$) that cannot be raised (lowered) by SMOC. For spherically symmetric systems  (e.g., atoms) all shells  contain an odd number of states, $2l+1=1,3,5\dots$ and have maximum (minimum) $m_l$ values, thus the odd $N$ cyclic and spherically symmetric cases are highly analogous. (b) In contrast, for even $N$  all states  couple to a state with equal total angular momentum about $z$, $j=k+\sigma$, e.g., $\ket{L;\uparrow}$ couples to $\ket{1-L;\downarrow}$. } 
	\label{fig:ring_v_ladder}
\end{figure}

For even $N$ the solutions of $L=(N-1)/2$ are half--odd-integers. However, if, instead, one defines $L=N/2$  for even $N$ and applies the selection rules [Eqs. (\ref{sel_rule})] one again finds that the $H_\textrm{SO}$ is given by Eq. (\ref{eq:Hodd}), but now  
$\lambda^z_0=\lambda^z_{L}=0$ by Eq. (\ref{eq:lz}). However, in the even case no $\lambda_j^\pm$ vanish by symmetry.
Thus, there are fundamental differences between odd- and even-membered rings, illustrated in Fig. \ref{fig:ring_v_ladder}. These are direct consequences of the modular addition, onto the interval $(-N/2,N/2]$, of angular momentum implicit in Eq. (\ref{eq:lz}).

{In orbitally degenerate systems, such as graphene \cite{Castro_Neto} and some transition metal oxides \cite{Nagaosa}, it is common to represent the orbital degeneracy via a pseudospin degree of freedom. Our results demonstrate that in molecular systems with strong SOC this may be problematic. For example, in $\mathcal{C}_2$ symmetric systems pairs of degenerate orbitals are fundamentally bosonic.}

\begin{figure*}
	\begin{center}
		{\bf a} \hspace{0.6\columnwidth}\hspace{0.4cm}\hspace{-5pt}
		{\bf b} \hspace{0.6\columnwidth}\hspace{0.4cm}\hspace{-5pt}
		{\bf c} \hspace*{0.6\columnwidth} 
		\\
		\includegraphics[width=0.6\columnwidth]{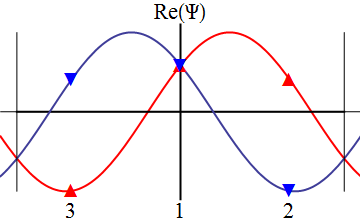}
		\hspace{0.4cm}
		\includegraphics[width=0.6\columnwidth]{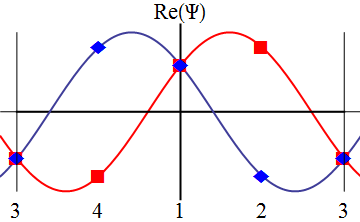}
		\hspace{0.4cm}
		\includegraphics[width=0.6\columnwidth]{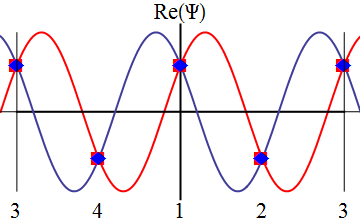}
	\end{center}
	\caption{
		Illustration of the aliasing of the maximum and minimum angular momentum states in even-fold symmetric molecules. Lines show the  molecular orbital angular momentum states defined in Eq. (\ref{Eq:Fourier}) in the $N\rightarrow\infty$ limit. We plot snapshots of the real part of the wavefunction as it evolves under the trivial Schr\"odinger time evolution. Here we show the wavefunctions for $\omega_k t=0.1$, where $\hbar\omega_k$ is the energy of the state $\ket{k}$. 
		(a) For three sites the maximum ($k=+1$, red) and minimum ($k=-1$, blue) angular momentum states are distinguishable when sampled on the three sites (marked on  the abscissa; values sampled are marked by triangles). 
		(b) On four sites the $k=+1$ (red) and  $k=-1$ (blue) angular momentum states remain distinguishable when sampled on the four sites (values sampled are marked by diamonds/squares).
		(c) However, the maximum ($k=+2$, red) and minimum ($k=-2$, blue) angular momentum states are \textit{indistinguishable} when sampled on the four sites (values sampled are marked by diamonds/squares); i.e., $\ket{+2}\equiv\ket{-2}$.
		For simplicity the phase factors, $\eta_k$, are not included in the figure, but, clearly, an overall phase factor cannot remove the aliasing. Animations of the full time evolution, shown in the Supplementary Information \cite{SI}, underscore that this argument holds at all times.
	} 
	\label{Fig:alias}
\end{figure*}

The differences between odd- and even-membered rings can be understood by examining the character table (\ref{tab:odd}). 
For even $N$ the Born--von K\'arm\'an boundary conditions of the ring imply that a single state  instantiates both the maximal and minimal molecular orbital angular momentum, $\ket{L}\equiv\ket{-L}$. 
In the language of signal processing, $\ket{L}$ and $\ket{-L}$ are aliases, see Fig. \ref{Fig:alias}.
 Hence $\ket{L;\uparrow}$ and 
$\ket{1-L;\downarrow}\in \overline E_{(N-1)/2}$; similarly $\ket{L;\downarrow}$ and $\ket{L-1;\uparrow}\in \overline E_{(1-N)/2}$. 
That is, there is always more than one state with the maximal (minimal) \emph{total} angular momentum, $j=k+\sigma$, and SMOC couples these  states. This  is highly analogous to umklapp scattering in  crystals.

{Indeed, an intuitively simple way to think about even-$N$ molecules is to take literally the statement that $L=N/2$ (e.g.,  $\mathcal{C}_2$ molecules have $L=1$), but remember that the $\ket{L}$ and $\ket{-L}$ states are identical, Figs. \ref{fig:ring_v_ladder}b and \ref{Fig:alias}. This gives a simple interpretation of why $\lambda^z_{L}=0$: because the state is both $\ket{L}$ and $\ket{-L}$ and thus `on average' $\hat L^z\ket{L}=\hat L^z\ket{-L}=0$. }


In contrast for odd $N$  different states  instantiate the maximal ($\ket{L;\uparrow}$) and minimal ($\ket{-L;\downarrow}$) \emph{total} angular momenta. Both of these states transform as $A_{N/2}$ and they form a Kramers doublet. Therefore,  time reversal symmetric terms in the Hamiltonian (such as SOC) cannot cause an interaction between $\ket{L;\uparrow}$ and $\ket{-L;\downarrow}$: this would lift their degeneracy, violating Kramers' theorem [Eq. (\ref{eq:Krammer})]. Thus the combination of ${\cal C}_N$ symmetry and time reversal symmetry leads directly to the close analogy with atomic SOC in the odd $N$ case.

{In the continuum limit ($N\rightarrow\infty$) the distinction between even and odd $N$ must vanish. This is apparent from previous solutions of problems described by this symmetry \cite{Hartmann}.}

	In real space the SMOC takes  the same form for both odd and even $N$: 
\begin{eqnarray}
H_{\mathrm{SO}} &=&\sum_{r\ne s,\alpha\beta}i{\bm \lambda}_{rs}\cdot{\bm \sigma}_{\alpha\beta}\hat a_{r\alpha}^\dagger \hat a_{s\beta}, \label{Eq:SOC_real_space}
\end{eqnarray}
where ${\bm \lambda}_{rs}=(\lambda^x_{rs},\lambda^y_{rs},\lambda^z_{rs})$,
\begin{subequations}
	\begin{eqnarray}
	\lambda^x_{rs} &=& \frac{1}{N}\sum_{j=1/2}^{L-1/2} [i\lambda^\pm_{j}e^{i\phi(r+s)/2} + c.c.] \sin\left[ \phi j(r-s) \right], 
	\\ 
	\lambda^y_{rs} &=& \frac{1}{N}\sum_{j=1/2}^{L-1/2} [\lambda^\pm_{j}e^{i\phi(r+s)/2} + c.c.]  \sin\left[ \phi j(r-s) \right], 
	\hspace{0.65cm}\\
	\lambda^z_{rs}
	&=& \frac{2}{N}\sum_{k=1}^{L}\lambda^z_{k}  \sin\left[ \phi k(r-s) \right],
	\end{eqnarray}
	\label{lxyz}
\end{subequations}
and $\hat a_{r\sigma}=\frac1{\sqrt{N}}\sum_ke^{i\phi kr}\eta_k\hat c_{k\sigma}$. Thus, ${\bm \lambda}_{rs}$ is a real vector even for complex $\lambda^\pm_{j}$.

\section{First principles calculations} 

The above arguments, based on symmetry considerations, only show that   SMOC is allowed. Therefore, it is natural to ask how large this effect is in real materials.
Most of the multi-nuclear complexes synthesized to date  with strong intermolecular coupling have not included heavy atoms. \Mo is a typical example \cite{LlusarReview,LlusarJACS}. In the absence of SOC its low-energy electronic structure is described by three Wannier orbitals per spin per molecule \cite{JackoPRB15}.
In the one-component (scalar) relativistic formalism the one finds a tight-binding model:
\begin{eqnarray}
H_1=\sum_{rs\sigma}t_{rs}^{(1)}\hat a_{r\sigma}^\dagger \hat a_{s\sigma}, \label{Eq:H1}
\end{eqnarray}
where $t_{rs}^{(1)}=\langle \psi_r|H|\psi_s\rangle$ is the hopping integral between  Wannier orbitals $\ket{\psi_r}$ and $\ket{\psi_s}$. A good model of the full density functional theory band structure can be achieved with only three hopping integrals: $t_c=60$~meV, intramolecular hopping; $t=47$~meV intermolecular hopping in the basal plane between  a single Wannier orbial on each molecule; and $t_z=41$~meV intermolecular hopping along the crystallographic $c$-axis from a Wannier orbital to the equivalent orbital translated in the $z$ direction  \cite{JackoPRB15}. 
Note that the hopping between any pair of Wannier orbitals within the same molecule is equivalent, consistent with the molecule's $\mathcal{C}_3$ symmetry.

\begin{figure}
	\begin{center}
		\includegraphics[width=\columnwidth]{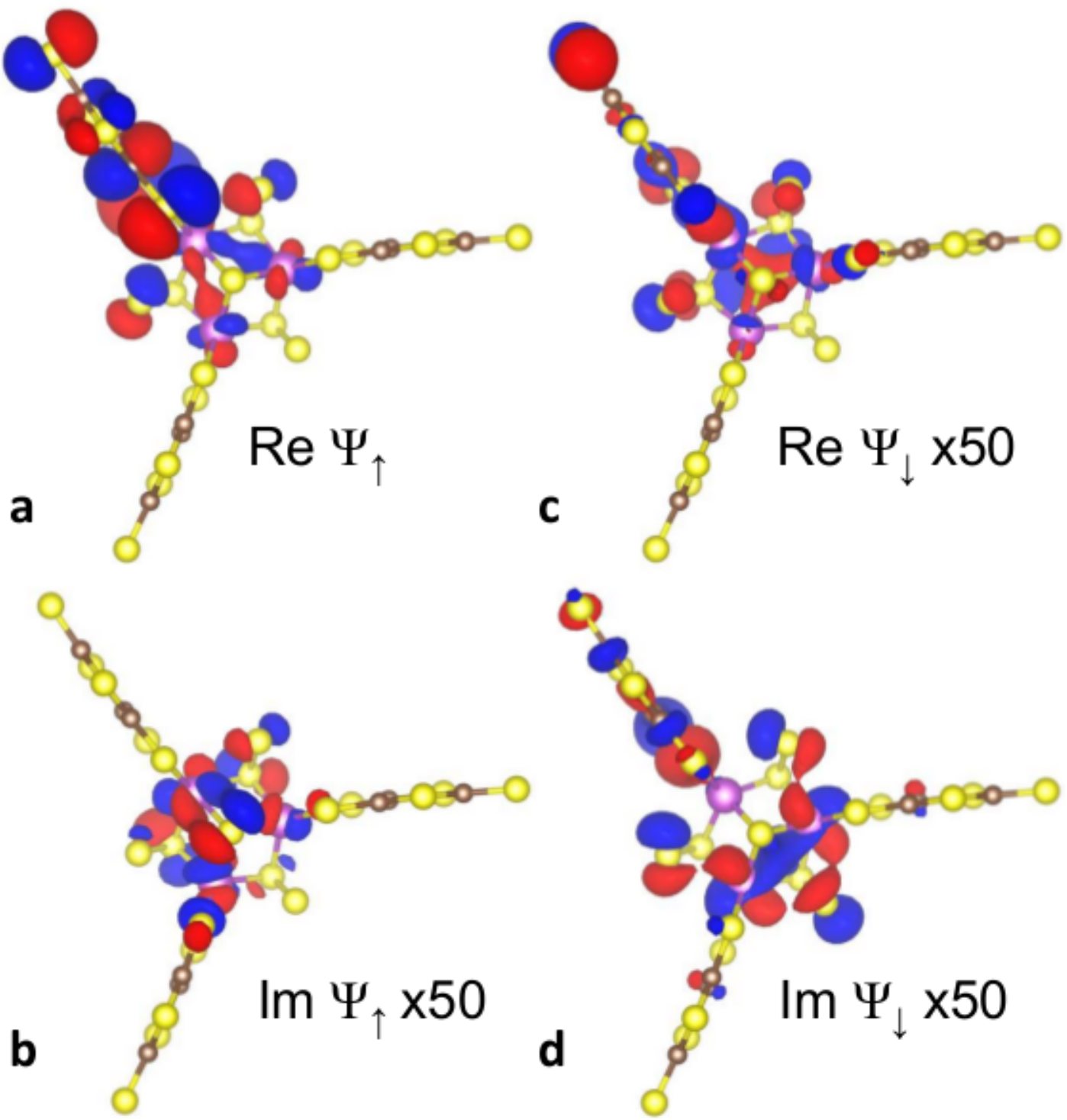}
	\end{center}
	\caption{The low energy physics of a single \Mo molecule can be understood in terms of six Wannier spinors (three Kramers pairs). The large components of one are shown above; the others are related by the $\tilde{\cal C}_3$ and/or time reversal symmetry. The four panels display the (a) real and (b) imaginary parts of the spin-up large component and the (c) real and (d) imaginary parts of the spin-down large component. Note that the isosurface  in panel (a) corresponds to a contour value fifty times smaller than those in panels (b-d). } 
	\label{fig:DFT}
\end{figure}

We solved the four-component Dirac-Kohn-Sham equation in an all-electron full-potential local orbital basis using the FPLO package \cite{koepernik99,Eschrig}.
The density was converged on an $(8 \times 8 \times 8)$ $k$-mesh using the  Perdew-Burke-Ernzerhof exchange-correlation functional \cite{perdew96}.
Localized Wannier \cite{Marzari} spinors were constructed from the twelve  bands closest to the Fermi energy, corresponding to six spinors (three Kramers pairs) per molecule. 
We calculated the overlaps between Wannier spinors (Fig. \ref{fig:DFT}) constructed from the solution of the  four-component Dirac-Kohn-Sham equation within the same molecule to construct a first principles single particle effective low-energy Hamiltonian.

The effective Hamiltonian in the four-component formalism is
\begin{eqnarray}
H_4&=&\sum_{rs\alpha\beta}t_{rs\alpha\beta}^{(4)}\hat a_{r\alpha}^\dagger \hat a_{s\beta},
\label{Eq:H4}
\end{eqnarray}
where 
$t_{rs\alpha\beta}=\langle \Psi_r^{(\alpha)}|H|\Psi_s^{(\beta)}\rangle$ is the hopping integral between the $\alpha$th component of the Wannier spinor $\ket{\Psi_r}$ and the $\beta$th component of  $\ket{\Psi_s}$. SU(2) invariance implies that $t_{rs\alpha\beta}^{(4)}=t_{rs}^{(4)}\delta_{\alpha\beta}+i{\bm \lambda}_{rs}\cdot{\bm \sigma}_{\alpha\beta}$. We find that $t_{rs}^{(4)}=t_{rs}^{(1)}$ for all hopping integrals investigated (all differences are $\ll1$~meV). The intramolecular SOC is given by ${\bm \lambda}_{12}=\lambda_0(-0.35,0.21,0.58)$,  ${\bm \lambda}_{23}=\lambda_0(-0.003,-0.42,0.58)$,  ${\bm \lambda}_{31}=\lambda_0(0.36,0.21,0.58)$ where we have numbered the three Wannier spinors on each molecule from one to three. Note that $\lambda_{rs}^z$ is the same for all pairs of Wannier spinors, but $\lambda_{rs}^x$ and $\lambda_{rs}^y$ vary significantly. This is precisely as predicted by Eqs. (\ref{lxyz}) with $\lambda^z_1=\lambda_0$ and $\lambda^\pm_{1/2}=0.72\lambda_0$. In the spherically symmetric case $\lambda^\pm_{1/2}/\lambda_z=\sqrt2$  for $L=1$ \footnote{This factor of $\sqrt2$ is simply  the relevant prefactor $\sqrt{(j\mp m)(j\pm m+1)}$ for spherically symmetric angular momentum ladder operators \cite{Sakurai}.}, so this corresponds to a significant anisotropy ($\sqrt2/0.72=1.96$).     Despite the relatively small atomic numbers of the constituent atoms the SMOC in \Mo is significant: $\lambda_0=0.1t=4.91$~meV, where $t$ is the largest intermolecular hopping integral. 

{Note that the tight-binding model, Eq. (\ref{Eq:H4}) contains, only one orbital per site. Thus, atomic transitions are integrated out of the tight-binding model and only the SMOC remains.
}

The Wannier spinor  (Fig. \ref{fig:DFT}) has significant  weight on the Mo atoms in the core and S atoms in the dmit ligands. This suggests substituting either, or both, of these for heavier atoms, e.g., W or Se, could significantly increase the relative strength of the SMOC (cf. Eq. (\ref{Pauli})), leading to a range of possible experimental avenues to engineer materials with exotic phases that require strong SOC.
To investigate the effects of heavier metals we considered 
W$_3$O(CCH$_3$)(O$_2$CCH$_3$)$_6$(H$_2$O)$_3$, which has a very similar electronic structure to \Mo \cite{Cotton}. 

However, the hopping between W$_3$O(CCH$_3$)(O$_2$CCH$_3$)$_6$(H$_2$O)$_3$ complexes is much weaker than that between \Mo complexes; thus the band structure based approach, employed for \Mo, is impractical. We therefore
calculated the electronic structure of a single complex both with and without SOC. These calculations were performed in a triple zeta plus polarization basis of Slater orbitals with the B3LYP functional \cite{B3LYP}  using the ADF package \cite{ADF}. 
The energies of the frontier orbitals in the one-component calculations were fit to Eq. (\ref{Eq:H1}), yielding an intramolecular hopping $t_c=174$~meV.  We then fit the corresponding molecular orbital energies in the four-component calculation to Eq. (\ref{Eq:H4}) with the SOC given by Eq. (\ref{Eq:SOC_real_space}). 
Again the SOC displays significant anisotropy, however in this complex the largest SOC constant $\lambda^{\pm}_{1/2}= 1.81t_c=315$~meV. Thus, like  the iridates  \cite{Kee,BalentsARCMP}, W$_3$O(CCH$_3$)(O$_2$CCH$_3$)$_6$(H$_2$O)$_3$ is in the strong SOC regime.

{We stress that the increase in the SMOC on moving from a Mo complex to a W complex does not imply that the SMOC is just a linear combination of atomic $\bm L\cdot\bm S$ terms. However, the potential, $V(\bm r)$ [cf. Eq. (\ref{Pauli})], is just a linear combination of atomic potentials and generically one expects that its gradient will be larger in systems composed of heavier atoms. }

{It is therefore natural to ask what ingredients lead to large SMOC. Four factors can be identified readily from the analysis above: 
\begin{enumerate}[(i)]
	\item The relevant molecular orbitals should have significant weight near the nuclei to ensure large expectation values for $\bm K$ in a given orbital.
	\item Large atomic number, $Z$ will result in larger $V(\bm r)$ and hence large SMOC. This is demonstrated by the above calculation.
	\item  As the nuclear potential varies most rapidly closest to the nuclei the heavy atoms should be close together to maximize  $\bm{\nabla}V(\bm r)$.
	\item SMOC is strongest for electrons  with large instantaneous  momenta. A semiclassical estimate of this can be made from the group velocity in the continuum limit. For nearest neighbor intramolecular hopping only
	this	 yields 
	$p={am_et_c}\sin(k\phi)/{\hbar}$,
	where $a$ is the distance between the centers of neighboring Wannier orbitals and $m_e$ is the mass of the electron. 
	The linear dependence of $p$ on $a$ is likely to be swamped by the rapid suppression of $t_c$ as $a$ increases. So the prefactor is likely to be largest if the Wanniers are close to one another.
	This is maximized for $k=\pm N/4$. However, these momenta are only realized in `anti-aromatic' compounds where $N=4n$ for integer $n$, suggesting that such molecules when close to half-filling  should have the largest SMOC.
\end{enumerate}
}

\section{Interplay of molecular spin-orbital coupling and electronic correlations}

\subsection{Spin-1/2 systems}

To examine the effects of  SMOC in cyclic molecules with strong electronic correlations we  analyzed the simplest example: the \tJ model for $\mathcal{C}_2$ molecules with two orbitals per molecule {at quarter filling (or equivalently three quarters filling due to particle hole symmetry)}. The Hamiltonian describing the $j$th molecule is
\begin{eqnarray}
H_{tJ}&=& P_0\left[\sum_{\sigma}\left(t_c\hat{a}^\dagger_{j1\sigma}\hat{a}_{j2\sigma}
+i\lambda \hat a^\dagger_{j1\overline\sigma}\hat{a}_{j2\sigma} + H.c.\right)
\notag\right.\\&&\left.\hspace{1cm}
+J_c\left( \hat{\bm S}_{j1}\cdot\hat{\bm S}_{j2} -\frac{\hat n_{j1}\hat n_{j2}}{4} \right) 
\right]P_0,
\end{eqnarray}
where $\hat{\bm S}_{j\mu}$ ($\hat n_{j\mu}$) is the spin (number) operator for the $\mu$th orbital on the $j$th molecule and $P_0$ projects out states that contain empty orbitals.  The $\mathcal{C}_2$ symmetry of the molecule implies that all $\lambda^z_k=0$  and $\lambda=-2i\lambda_{1/2}^{\pm}\in\mathbb{R}$. {This means that the SMOC only couples the $x$ components of the spin and orbital degrees of freedom, cf. Eqs. (\ref{lxyz}).}

\begin{figure*}
	\includegraphics[width=\columnwidth]{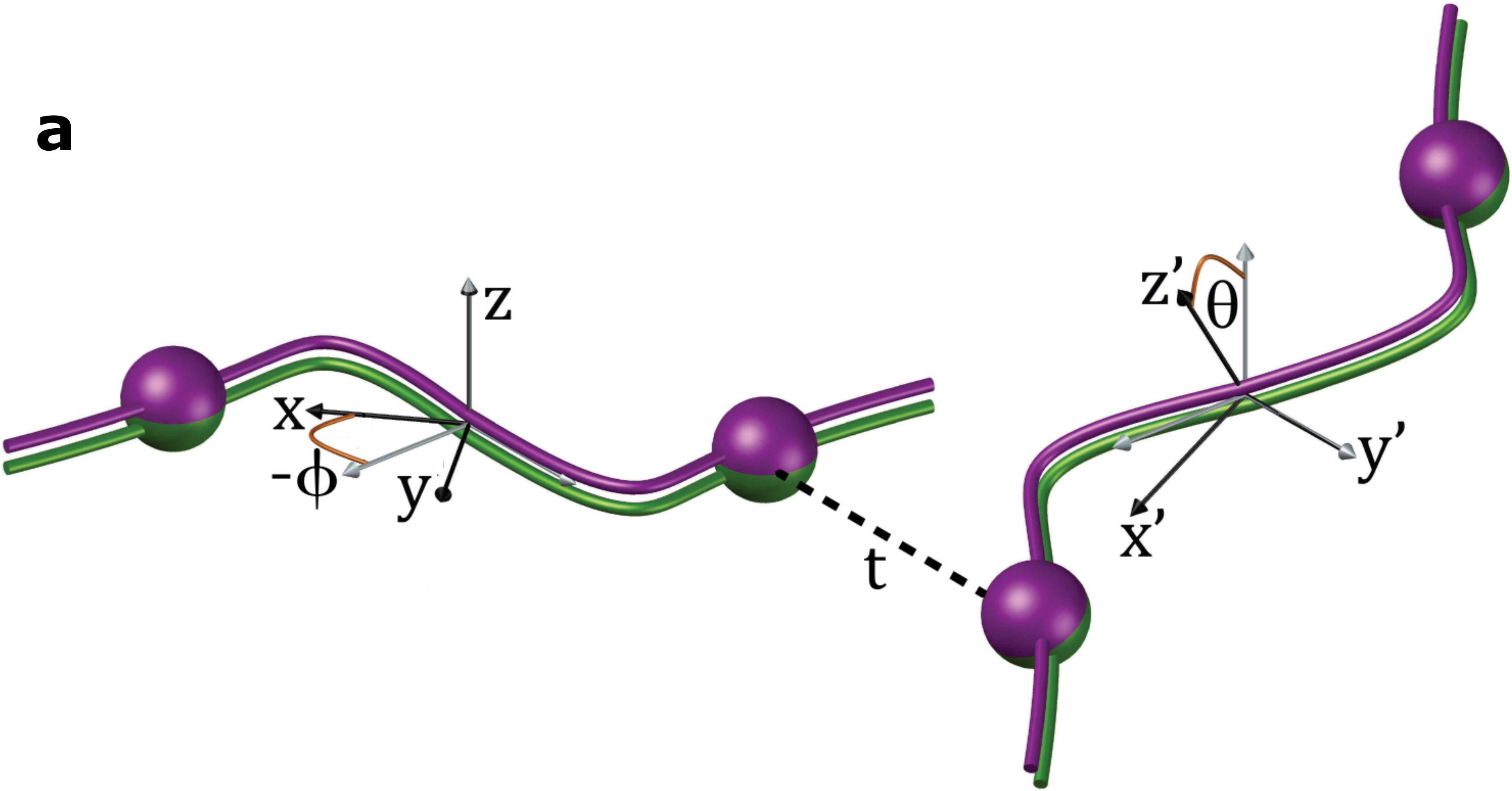}   
	\includegraphics[width=0.67\columnwidth]{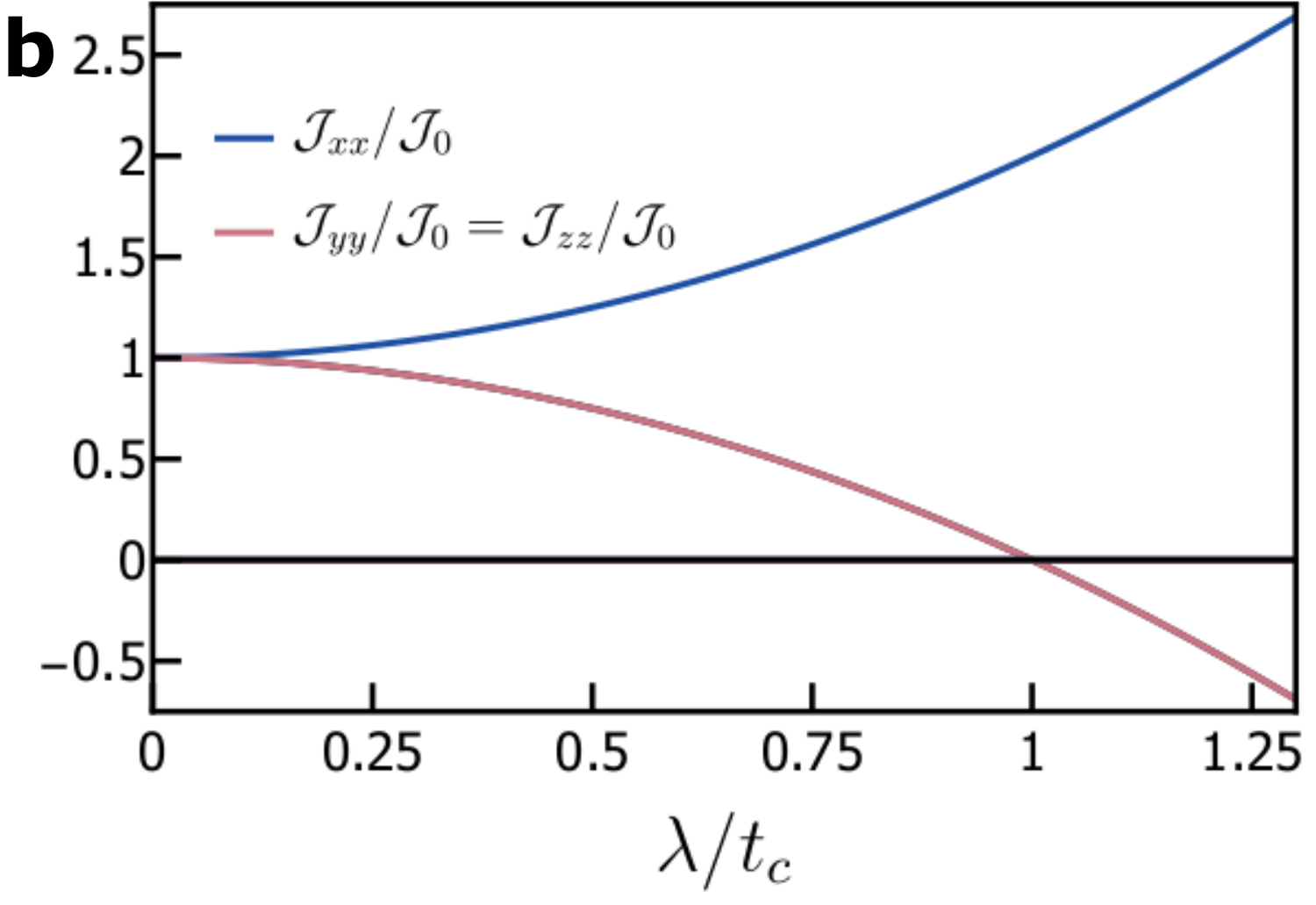} 	
	\\
	\includegraphics[width=0.67\columnwidth]{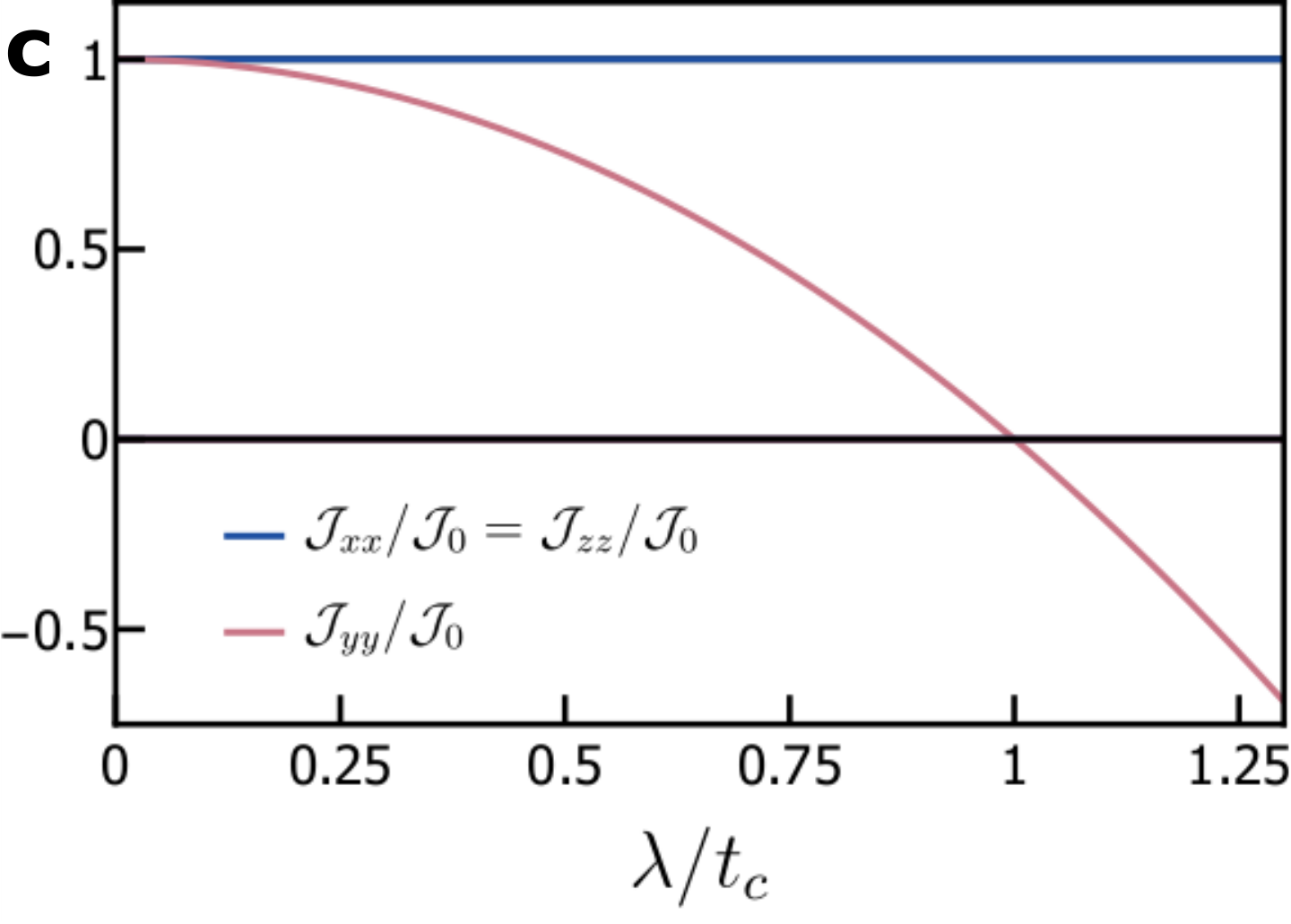} 
	\includegraphics[width=0.67\columnwidth]{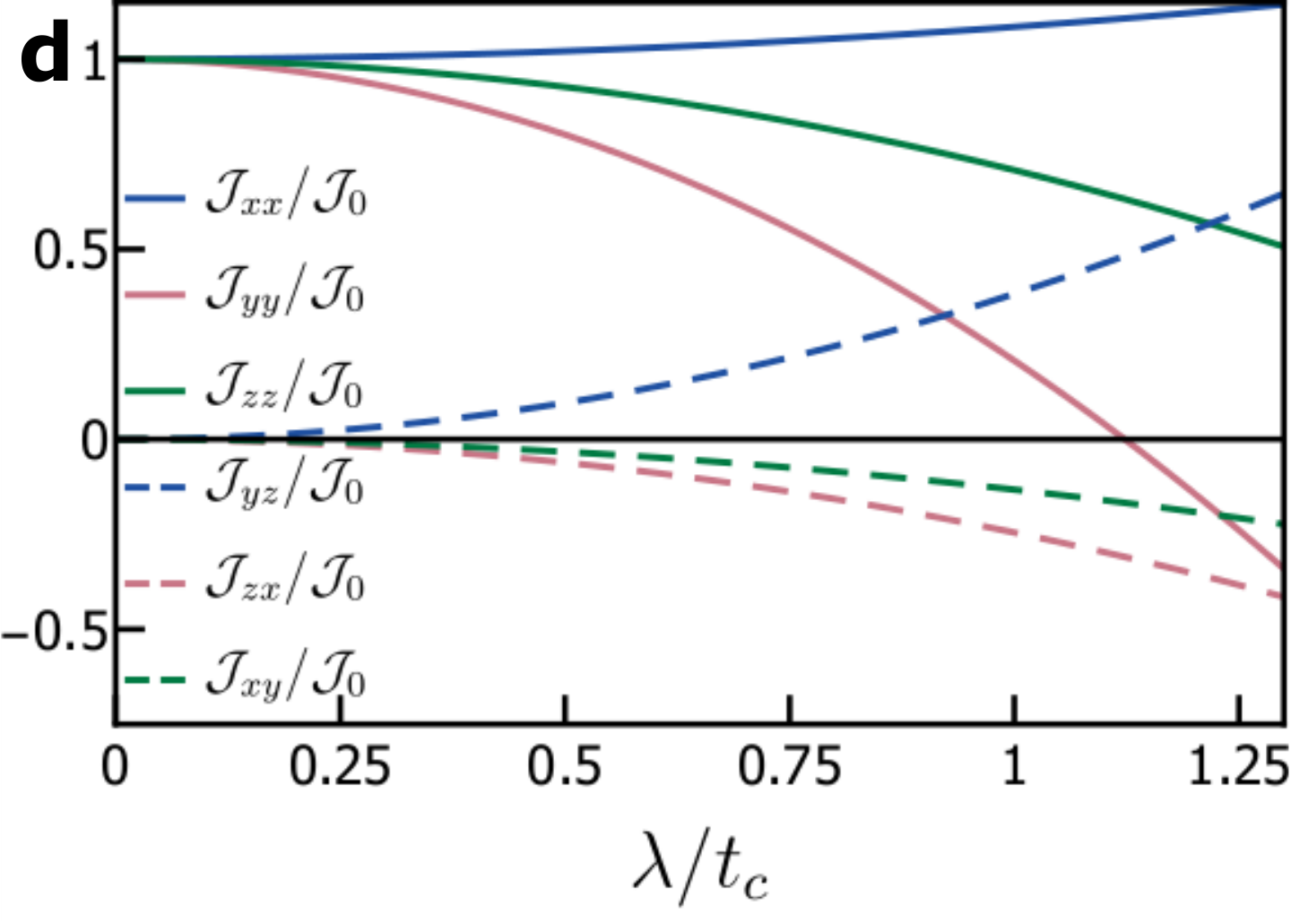} 
	\includegraphics[width=0.67\columnwidth]{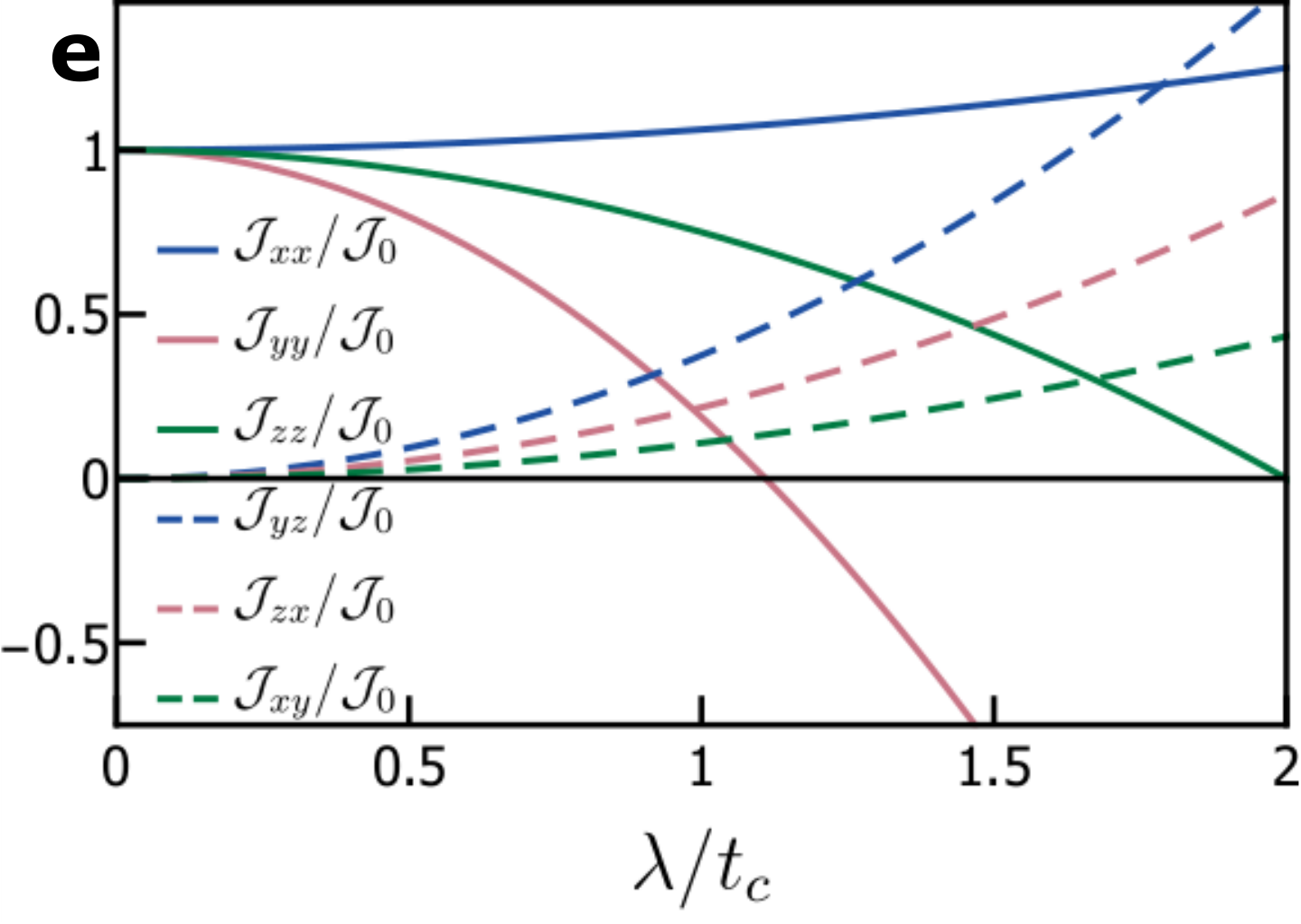}
	\caption{The exchange anisotropies vary significantly in molecular materials with different packing motifs. 
		(a) Sketch of a pair of nearest neighbors in the \tJ model for $\mathcal{C}_2$ molecules. Spheres indicate the Wannier orbitals and the curves connecting them show the molecular symmetry. The nearest neighbor intermolecular hopping, $t$, is marked. The local $x$ and $y$ axes are uniquely determined by the SMOC via the phase convention chosen  in Eq. (\ref{Eq:Fourier}).  
		Thus we parametrize the packing motif by the angles between the local axes on neighboring molecules: $\theta$ ($\phi$) is the relative rotation about the $y$ ($z$) axes; the effective Hamiltonian is independent of rotations about the $x$-axes. The local coordinate system is shown in black and the angles are marked relative to the gray axes, which are point in the same directions on both molecules. As we only consider pairwise interactions we write the effective Hamiltonian in the local coordinates of the $i$th molecule, cf. Eq. (\ref{Eq:anisoHeis}). 
		(b) Parallel stacking ($\theta=\phi=0$) leads to Ising anisotropy. 
		(c) Perpendicular packing ($\theta=\pi/2$, $\phi=0$) gives XY anisotropy. 
		(d,e) More complicated packing leads to {lower symmetry exchange Hamiltonians} [here we plot (d) $\theta=\phi=1$ and (e) and $\phi=-\theta=2\pi/3$]. 
		In all plots $J=0$ and ${\cal J}_0 = {t^2t_c^2J_c}/\{2[t_c^2+\lambda^2][2(t_c^2+\lambda^2)-J_c\sqrt{t_c^2+\lambda^2}]\}$. 
		Analytical expressions for ${\cal J}_{\alpha\beta}$ and ${\bm D}^\pm$ are given in Appendix \ref{magInt}.
	}
	\label{fig:J}
\end{figure*}

If neighboring molecules are related by  inversion  then $\lambda$ is the same on both molecules. However, if they are related by a  $\pi$ rotation about the $z$-axis,  $\lambda$ must be of equal magnitude but opposite sign on the two molecules. For simplicity, we assume $\lambda$ has the same magnitude on all molecules  and  consider arbitrary orientations of the molecules. We  include  \tJ interactions between molecules:
$ P_0\sum_{\langle ij\mu\nu\rangle\sigma}[t(\hat{a}^\dagger_{i\mu\sigma}\hat{a}_{j\nu\sigma}
+ H.c.)+J( \hat{\bm S}_{i\mu}\cdot\hat{\bm S}_{j\nu} -{\hat n_{j1}\hat n_{j2}}/{4} )]P_0$,
where the angled brackets imply that the sum runs only over nearest neighbor orbitals, cf. Fig. \ref{fig:J}a.  We consider  a ground state with one hole per molecule and assume that we are in a parameter regime consistent with a bulk molecular Mott insulator \cite{JPCM,KK}.  
The effective interactions between neighboring molecules were evaluated analytically using the DiracQ package \cite{DiracQ}  in Mathematica.
To second order in $t$ (and hence first order in $J$)  one finds a low-energy effective Hamiltonian describing pseudospin-1/2 degrees  of freedom, $\hat{\bm{\mathcal{ S}}}_j=(\hat{\mathcal S}_j^x,\hat{\mathcal S}_j^y,\hat{\mathcal S}_j^z)$, on each molecule:
\begin{eqnarray}
\cal{H}_\textrm{eff}^\pm&=&\sum_{ij\alpha\beta}{\cal J}_{\alpha\beta}\hat{\mathcal S}_i^\alpha\hat{\mathcal S}_j^\beta 
+\sum_{ij}{\bm D}^\pm\cdot\hat{\bm{\mathcal S}}_i\times\hat{\bm{\mathcal S}}_j
+\varepsilon_0
,\label{Eq:anisoHeis}
\end{eqnarray}  
where  $\pm$ indicates the relative signs of $\lambda$ on the two molecules. 
The exchange, ${\cal J}_{\alpha\beta}$, and Dzyaloshinskii-Moriya coupling, ${\bm D}^\pm$,  are both strongly dependent on the relative orientation of the molecules.  ${\cal J}_{\alpha\beta}$ is highly anisotropic, cf. Fig. \ref{fig:J}, and independent of the relative signs of $\lambda$.

Hence, the  SMOC leads to anisotropic exchange interactions. Furthermore, the anisotropy is strongly dependent on the relative orientation of the molecules.
Thus it is possible to vary the exchange anisotropies between distinct pairs of molecules by arranging them in packing motifs with different angles between  the pairs, cf. Fig. \ref{fig:J}.  
This would open the way to providing new realizations of compass models, such as the Kitaev model \cite{Brink,Kitaev}. 
The inclusion of 5d metals opens up the possibility of reaching  large effective SMOCs ($\lambda>t_c$) in molecular crystals, as found in W$_3$O(CCH$_3$)(O$_2$CCH$_3$)$_6$(H$_2$O)$_3$.

\subsection{{Spin-one systems}}

If the molecules are half filled (two electrons in two orbitals) then one must use the full Hubbard model rather than the \tJ model. We also include a (ferromagnetic) direct exchange interaction, $J_F$, which is analogous to the atomic Hund's rule coupling and will play a crucial role in the analysis below. We have also analyzed other possible Coulombic interactions, and while these have some qualitative effects they are not  qualitatively important, so for simplicity we will not discuss them below.
Thus, we consider the extended Hubbard model. For the $j$th molecule the Hamiltonian is
\begin{eqnarray}
H_{xH}&=& \sum_{\sigma}\left(t_c\hat{a}^\dagger_{j1\sigma}\hat{a}_{j2\sigma}
+i\lambda \hat a^\dagger_{j1\overline\sigma}a_{j2\sigma} + H.c.\right)
\notag\\&&
-J_F \hat{\bm S}_{j1}\cdot\hat{\bm S}_{j2} +U\sum_{r=1}^2 \hat{n}_{jr\uparrow}\hat{n}_{jr\downarrow}, \label{HxH}
\end{eqnarray}
where $U$ is the effective Coulomb interaction between two electrons occupying the same Wannier orbital.

\begin{figure*}
	\includegraphics[height=0.47\columnwidth]{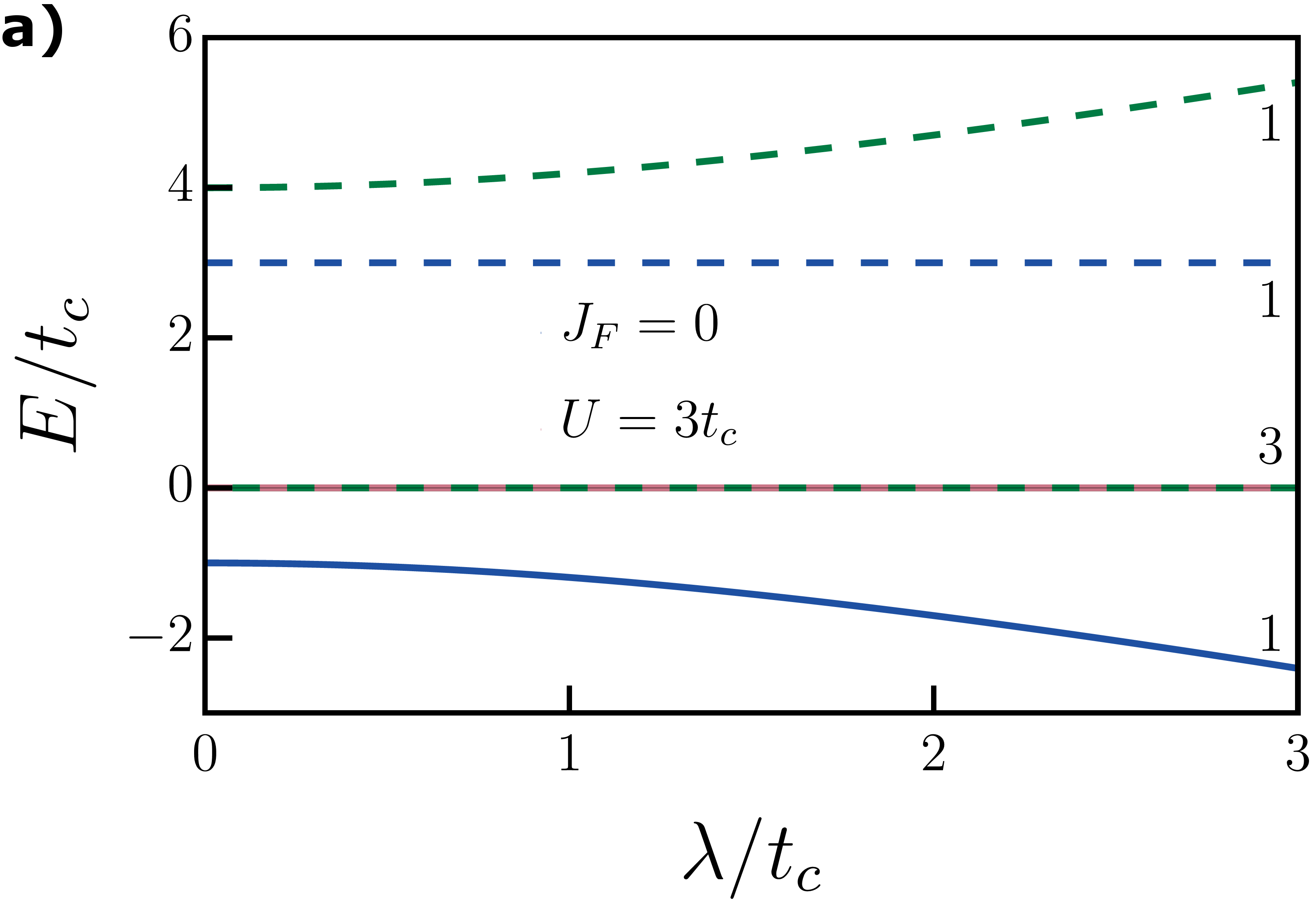}   
	\includegraphics[height=0.47\columnwidth]{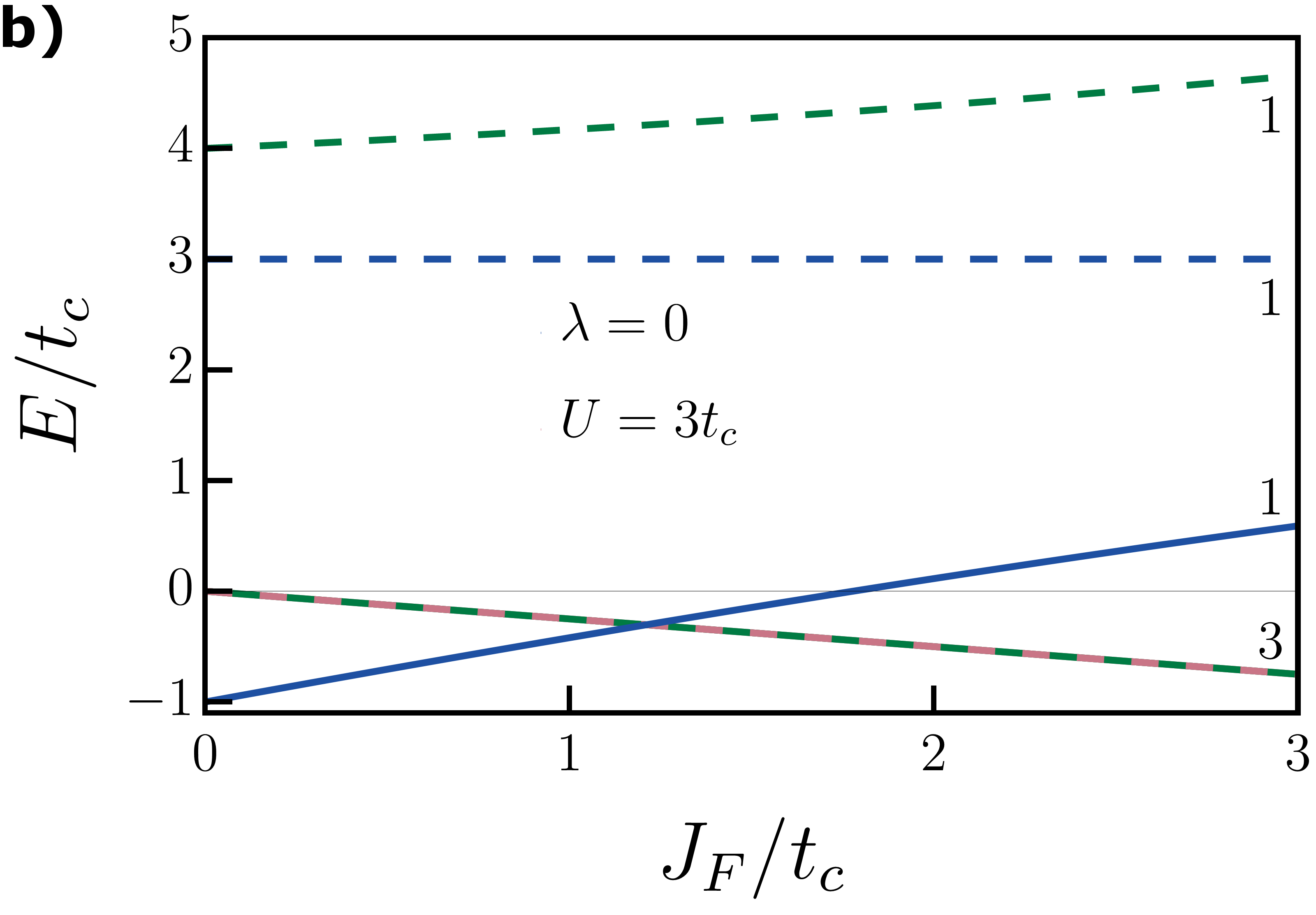} 
	\includegraphics[height=0.47\columnwidth]{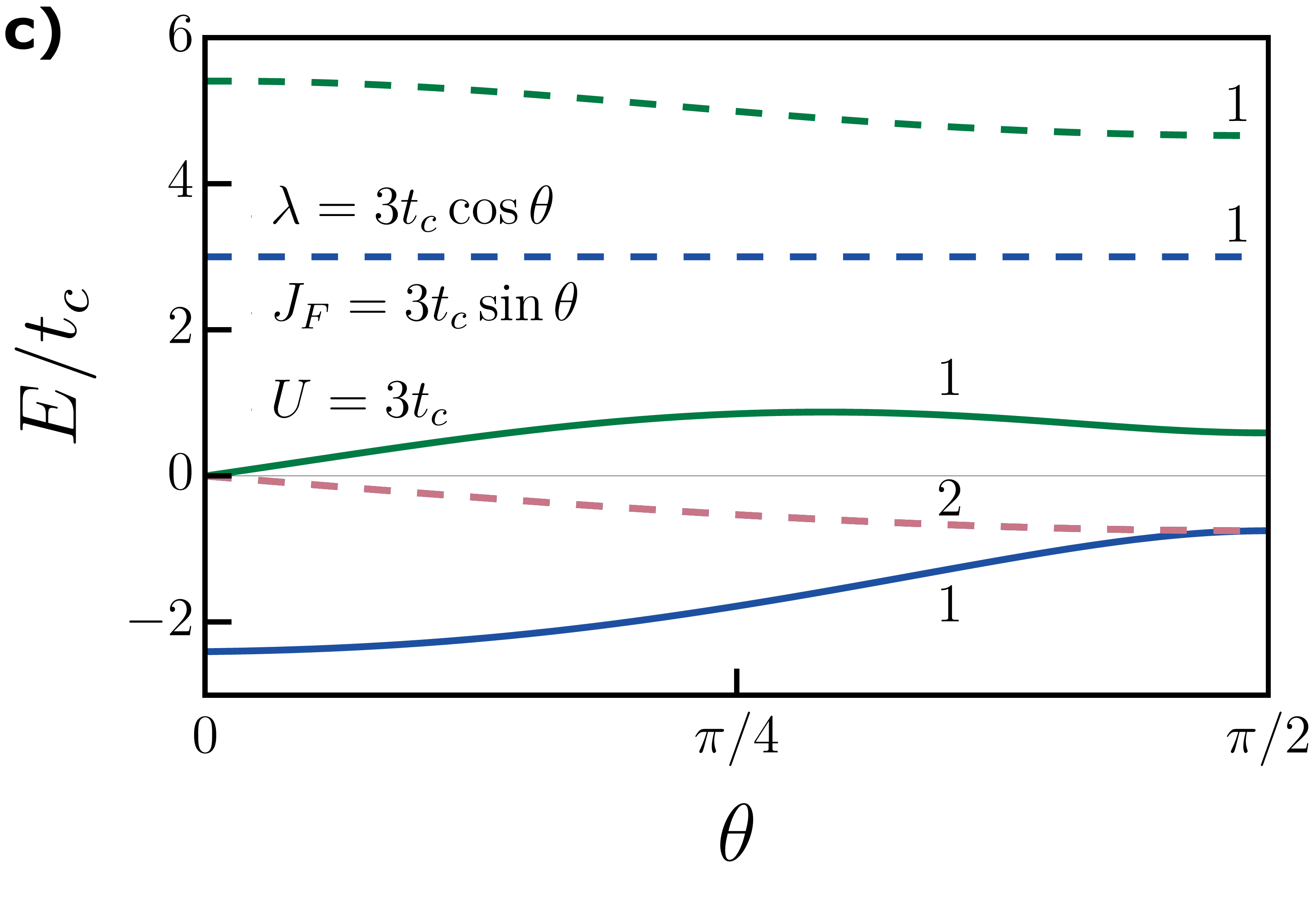} 	
	\\\vspace*{10pt}
	\includegraphics[width=0.67\columnwidth]{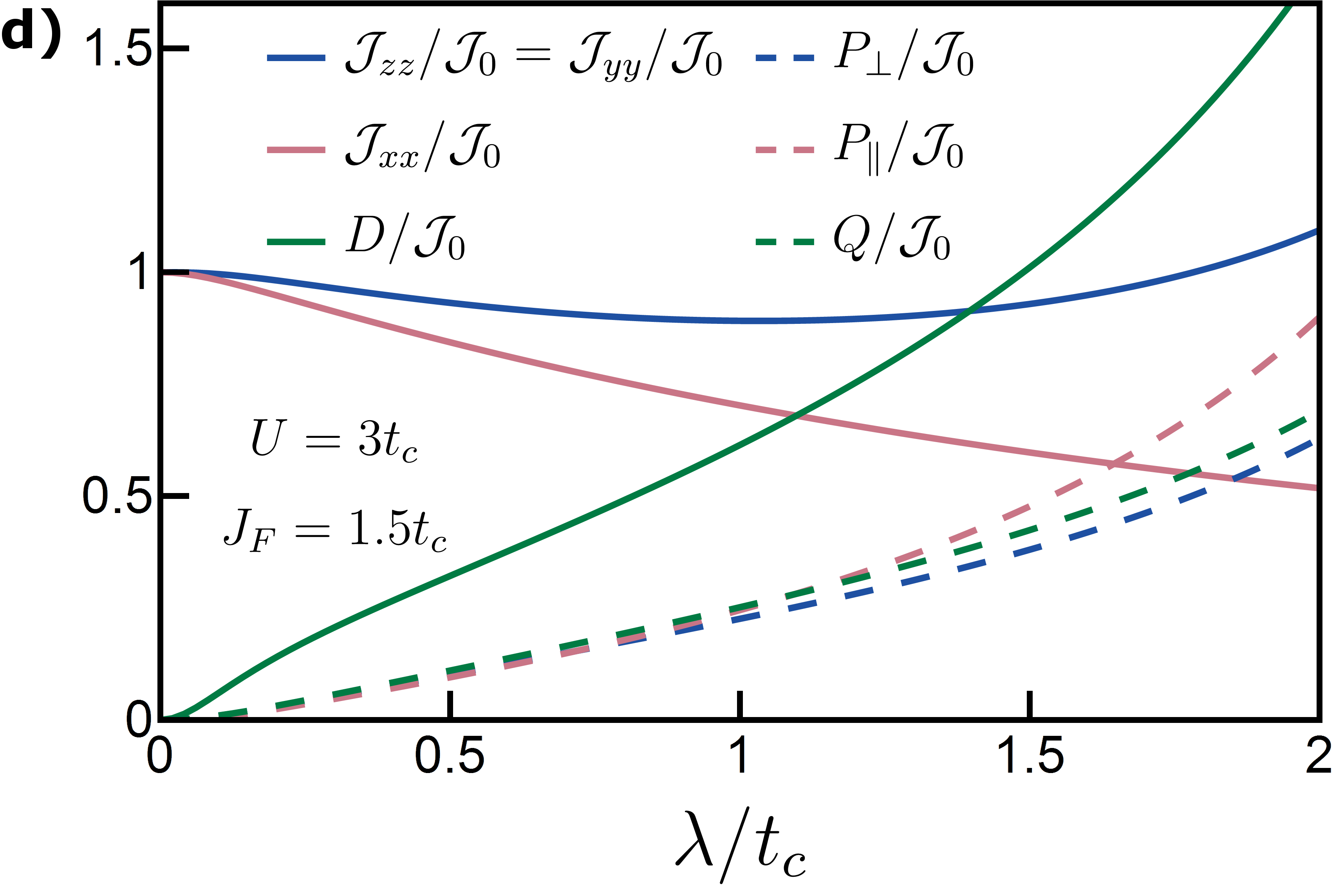} 
	\includegraphics[width=0.67\columnwidth]{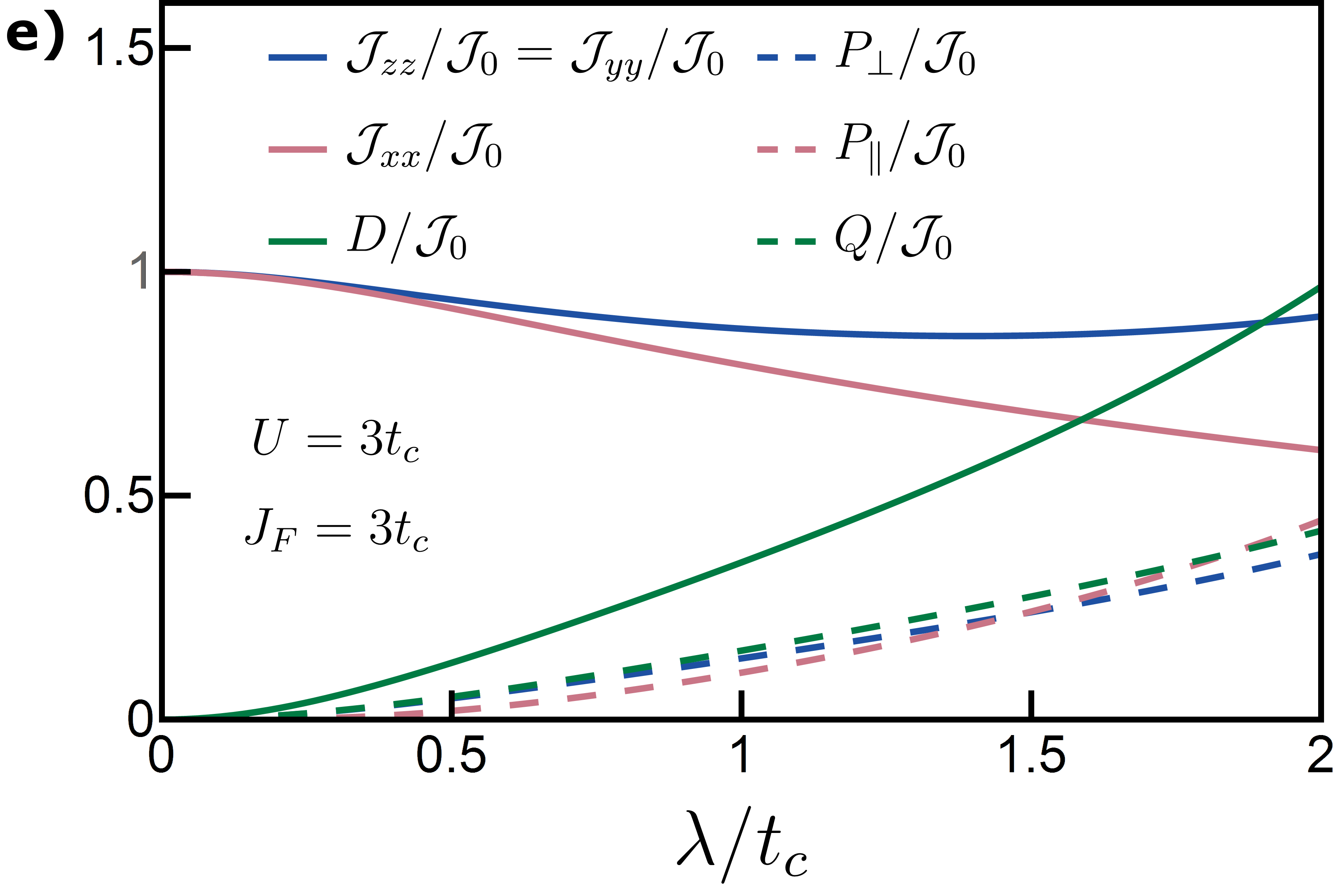} 
	\includegraphics[width=0.67\columnwidth]{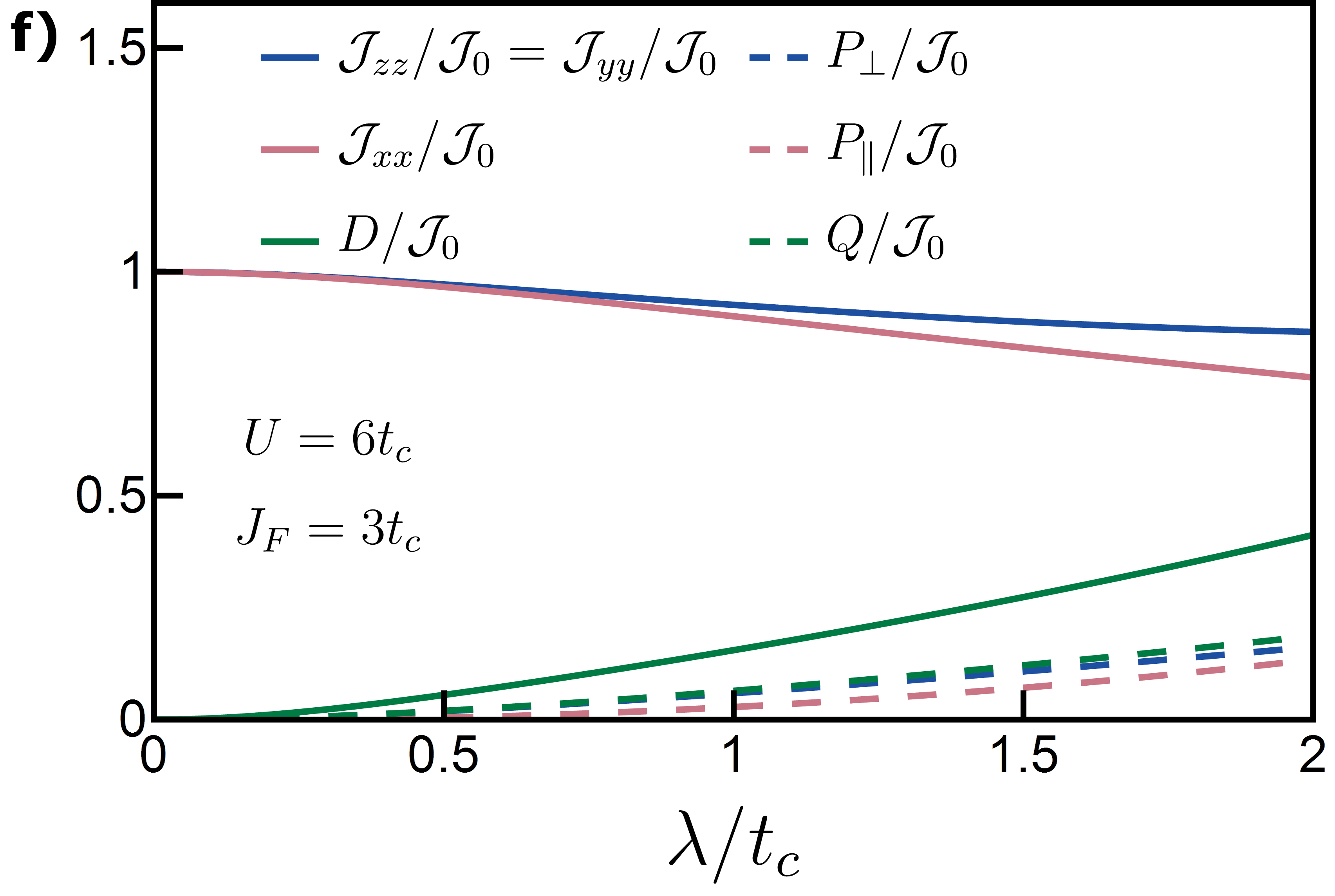} 
	\caption{{Effective spin-one model for half-filled $\mathcal{C}_2$ molecules. (a-c) Spectrum of a single $\mathcal{C}_2$ molecule with two electrons in two orbitals with (a) $J_F=0$, $\lambda\ne0$,  (b) $J_F\ne0$, $\lambda=0$, and (c)  $J_F\ne0$ and $\lambda\ne0$. Numbers on the right hand side of these panels label the degeneracies of the states. In  cases (a) and (b) the low-energy states are a singlet and a triplet. Only when both  $J_F$ and $\lambda$ are non-zero is the degeneracy of the triplet lifted, (panel c). (d-f) Parameters of the effective spin-one model [Eq. (\ref{Eq:spin1})]. As in the spin-1/2 case (Fig. \ref{fig:J}) SMOC causes large anisotropic interactions. These are most pronounced when the lowest energy spin-singlet excitation on a single molecule is at low energies}.
	}
	\label{fig:spin1}
\end{figure*}

In order to understand the effective interaction between neighboring molecules it is helpful to first understand the single molecule problem. For $\lambda=J_F=0$ the ground state is a (Coulon-Fischer) singlet and the first excited state is a triplet, with the remaining excited states having strong charge transfer character \cite{BenNotes}. Thus materials composed of molecules described by this parameter regime are magnetically inert.

Remarkably, turning on $\lambda$ at $J_F=0$ does \textit{not} change the degeneracy of this  spectrum, Fig. \ref{fig:spin1}a.
The degeneracies in the $\lambda=0$ case are usually understood  in terms of the $SU(2)\times SU(2)$ symmetry under rotations in spin-space. At non-zero $\lambda$ Hamiltonian (\ref{HxH}) retains an $SU(2)\times SU(2)$ symmetry under simultaneous lockstep rotations in spin- and orbital-space. Thus the degeneracies remain. {Note that, already at the single molecule
	level a $\mathcal{C}_2$ symmetric molecule is significantly  different from a $\mathcal{C}_3$ symmetric molecule at two thirds filling, where the ground state is a spin-triplet even for  $J_F=0$. But, in the $\mathcal{C}_3$ symmetric case this spin-one  manifold is split for any non-zero SMOC even when  $J_F=0$ \cite{Jaime}. }

Alternatively, at $\lambda=0$ the most significant effect of $J_F$ is to drive a level crossing between the Coulson-Fischer singlet and the triplet, Fig \ref{fig:spin1}b. When both $\lambda$ and $J_F$ are non-zero this crossing is avoided, Fig. \ref{fig:spin1}c, concomitant with strong mixing of the Coulson-Fischer singlet and one of the triplets. The $\mathcal{C}_2$ symmetry implies that there is only an $x$-component of the SMOC [cf. Eqs. (\ref{eq:Hodd}) and (\ref{HxH}), and Fig. \ref{fig:J}a] and hence  only the $S^x=0$ triplet mixes with singlet.  This has important consequences for the effective spin models that we discuss below.

We carry out the perturbation theory as above, with the appropriate perturbative coupling for the Hubbard model, i.e., $ \sum_{\langle ij\mu\nu\rangle\sigma}t(\hat{a}^\dagger_{i\mu\sigma}\hat{a}_{j\nu\sigma}
+ H.c.)$. While this calculation can be carried out exactly, we were unable to derive closed form expressions for the effective parameters as we were in the three quarters filled case. Given the greater parameter space of this problem we limit the discussion below to the inversion symmetric case. However, we do again find that in this problem the nature and the anisotropy of the Hamiltonian is again controlled by the molecular packing.

In the regime where the low energy part of the spectrum contains three  states per molecule, we find that the low-energy physics is described by a pseudospin-one model with the effective Hamiltonian
\begin{widetext}
\begin{eqnarray}
{\cal H}_\textrm{eff}^{\mathcal{C}_2}&=&\sum_{ij\alpha}{\cal J}_{\alpha\alpha}\hat{\mathcal S}_i^\alpha\hat{\mathcal S}_j^\alpha 
+\sum_{i}D \hat{\mathcal S}_i^x\hat{\mathcal S}_i^x
+\varepsilon_0
+Q \sum_{ij} (1-\hat{\mathcal S}_i^x\hat{\mathcal S}_i^x)\hat{\mathcal S}_j^x
+P_{\|}  \hat{\mathcal S}_i^x\hat{\mathcal S}_i^x \hat{\mathcal S}_j^x\hat{\mathcal S}_i^x
\notag\\&&
+P_{\perp} \left( \hat{\mathcal S}_i^y\hat{\mathcal S}_i^x \hat{\mathcal S}_j^y\hat{\mathcal S}_i^x
+\left[ \hat{\mathcal S}_i^y + i\hat{\mathcal S}_i^z\right] 
\left[ \hat{\mathcal S}_j^y - i\hat{\mathcal S}_j^z\right]  \left[\left(\hat{\mathcal S}_i^x + \hat{\mathcal S}_j^x\right)^2 - \hat{\mathcal S}_i^x - \hat{\mathcal S}_j^x - 1\right]
+H.c.
\right)
+\varepsilon_0.
\label{Eq:spin1}
\end{eqnarray}
The effective parameters of this model are plotted for various microscopic parameters in Fig. \ref{fig:spin1}d-f. For all microscopic parameters we find that ${\cal J}_{yy}={\cal J}_{zz}$. For $\lambda=0$ this model reduces to the isotropic spin-one Heisenberg model. Therefore, for example, if one considers a chain of such molecules the system would realize the Haldane phase -- a symmetry protect topological phase. At non-zero $\lambda$ the additional terms in the Hamiltonian pushes the system in different directions. The terms proportional to $P_\|$ and $P_\perp$ contain (some of) the biquadratic terms in the  Affleck-Kennedy-Lieb-Tasaki (AKLT) model \cite{AKLT} (as well as some additional terms) and therefore presumably stabilize the Haldane phase. In contrast the term proportional to $D$ favors a topologically trival phase where all the spins take the state $\hat{\mathcal S}_i^x=0$. As $D$ increases most rapidly with $\lambda$ one presumes that for large enough $\lambda$ this so-called D-phase is realized.

It is interesting to compare the effective Hamiltonian in Eq. (\ref{Eq:spin1}) with the effective spin-one model for four electrons (two holes) per molecule in a $\mathcal{C}_3$ symmetric molecule (a full analysis of this problem is reported elsewhere \cite{Jaime}). In the latter case $J_F$ plays a somewhat more subtle role, as it is not required to stabilize a triplet ground state in the absence of SOC, nevertheless both  $J_F$ and $\lambda$ must be non-zero for the effective model to acquire anisotropic exchange interactions in both cases. In  $\mathcal{C}_3$ molecules the SMOC is rather different from that in $\mathcal{C}_2$ molecules. In particular, the the $x$ and $y$ components are equal but, in general, different from the $z$ component of the SMOC in the $\mathcal{C}_3$ case, which leads to very different effective Hamiltonains. For a simple inversion symmetric coupling between neighboring orbitals (analogous to the $\mathcal{C}_2$ case above) one finds \cite{Jaime} that
\begin{eqnarray}
{\cal H}_\text{eff}^{\mathcal{C}_3}&=& 
\sum_{ij\alpha\beta}{\cal J}_{\alpha\beta} {\mathcal S}^\alpha_i {\mathcal S}^\alpha_j 
+ \sum_i \Big\{ D{\mathcal S}^z_i{\mathcal S}^z_i
+\big[ K_{\pm\pm}  {\mathcal S}_i^+{\mathcal S}_i^+
+    K_{z\pm}{\mathcal S}_i^z {\mathcal S}_i^x 
\label{eq:istr} 
+H.c. \big] \Big\}
+\varepsilon_0,
\label{eq:ab}
\end{eqnarray}
\end{widetext}
where ${\cal J}_{\alpha\beta}={\cal J}_{\beta\alpha}$.
This model is radically different from Eq. (\ref{Eq:spin1}): (i) the diagonal Heisenberg exchange terms are all different (${\cal J}_{zz}>{\cal J}_{yy}>{\cal J}_{xx}$); (ii) the intramolecular terms proportional to $K_{\pm\pm}$ and $K_{z\pm}$ are absent from Eq. (\ref{Eq:spin1}); (iii) in Eq. (\ref{eq:ab}) ${\cal J}_{xy}={\cal J}_{yz}=0$, but ${\cal J}_{zx}\ne0$, whereas there is no such off-diagonal exchange terms in Eq. (\ref{Eq:spin1}); and (iv) the higher-order terms proportional to $Q$, $P_\|$, and $P_\perp$ are absent from Eq. (\ref{eq:ab}). Interestingly, biquadratic exchange terms can be induced in the effective Hamiltonian for $\mathcal{C}_3$ molecules, for example if they are stacked so as to form a triangular tube, which breaks inversion symmetry. However, even in this case the effective Hamiltonian retains important differences from Eq. (\ref{Eq:spin1}) \cite{Jaime}.

Thus it is clear that the different forms of the SMOC for $\mathcal{C}_2$ and $\mathcal{C}_3$ symmetric  molecules result in very different effective magnetic interactions.

\section{Conclusions}

Thus, we have seen that in systems with cyclic symmetry the SOC is modified from the usual spherically symmetric case. {In particular, the  SMOC is not just inherited from the atomic scale, but an emergent property at the molecular scale. In cyclic molecules, decorated lattices and nanostructures}, the electronic spin couples to currents flowing around the molecule, rather than to intra-atomic angular momentum. For odd $N$ time reversal symmetry forbids umklapp-like spin-orbit scattering raising or lowering the molecular angular momentum  across the Brillouin zone boundary. However, for even $N$ all molecular angular momentum states can be raised and lowered -- this is a direct consequence of the the maximum and minimum molecular angular momenta being aliases for a single state. {Cyclic molecules provide an appealing context for understanding SMOC, as the interpretation in terms of angular momentum around the molecule is similar to the spherically symmetric case familiar from atomic physics. Nevertheless similar analyses can be carried out for molecules or nanostructures with arbitrary symmetry and in general will have a more complex interpretation. }

Density functional calculations demonstrate that the coupling of spin to molecular orbital angular momenta is large in suitable multi-nuclear organometallic complexes compared to the energy differences between frontier molecular orbitals. However, we stress that our results are not limited to these materials; and apply to all systems with appropriate cyclic symmetry.

We have discussed the consequences of this  SMOC for exchange anisotropy in materials with strong electronic correlations. These calculations demonstrate that together molecular packing and SMOC provide methods of controlling and engineering SOC Hamiltonians that are not available in traditional inorganic materials where the SOC arises from atomic processes. {Furthermore, we have shown that the symmetry of the molecule has a dramatic effect on the form of the effective magnetic interactions.} In fields as diverse as spintronics, organic light-emitting diodes, molecular qubits, and designing topological phases of matter  major problems could be solved if one had excellent control of SOC \cite{Manchon,Bihlmayer,CCR,Brink,Shiddiq,Freedman,BalentsARCMP,Kee}.  The ideas presented above have potential applications in all of these areas.

\section*{Acknowledgments}

We thank  Tom Stace and Xiuwen Zhou for helpful conversations. This work was supported by the Australian Research Council through grants FT130100161, DP130100757, DP160100060 and LE120100181. J.M. acknowledges 
financial support from (MAT2015-66128-R) MINECO/FEDER, UE. Density functional calculations were performed with resources from the National Computational Infrastructure (NCI), which is supported by the Australian Government.

\appendix

\section{Derivation of the selection rules  [Eqs. (\ref{sel_rule})]}\label{sect:rules}

$H_{\mathrm{SO}}$, like all Hamiltonian elements, belongs to the trivial representation $A_0$; it follows immediately from the $\tilde{\mathcal{C}}_N$ multiplication tables that 
\begin{eqnarray}
\bra{\overline{j}_\mu}H_{\mathrm{SO}}\ket{\overline{i}_\nu}=\Lambda_{j;\mu\nu}\delta_{ij},
\label{eq:consj}
\end{eqnarray}
where $\Lambda_{j;\mu\nu}$ is a constant. Thus, $H_{\mathrm{SO}}$ conserves $j$. 

Time reversal symmetry implies that $H_\textrm{SO}={\cal T}^{-1}H_\textrm{SO}{\cal T}$. For fermionic representations ${\cal T}^2\ketl{\overline{j}_\mu}=-\ketl{\overline{j}_\mu}$. Thus, the antiunitarity of ${\cal T}$ implies that
\begin{eqnarray}
\bra{\overline{j}_\mu}H_{\mathrm{SO}}{\cal T}\ket{\overline{i}_\nu} = -\bra{\overline{i}_\nu}H_{\mathrm{SO}}{\cal T}\ket{\overline{j}_\mu}.
\end{eqnarray}
Setting $\ketl{\overline{j}_\mu}=\ketl{\overline{i}_\nu}$ yields  Eq. (\ref{eq:Krammer}).

Further progress can be made by noting the explicit form of $H_{\mathrm{SO}}$, Eq. (\ref{Pauli}). In particular, $\bm \sigma$ acts only on the spin subspace whereas $\bm K
$ acts only on the molecular orbital subspace. 
$K^z$, $K^+$, and $K^-$ transform according to $A_0$, $E_1$, and $E_{-1}$ respectively for $N\geq3$. 
Thus,
\begin{eqnarray}
\Big\langle{\underline{k}_\mu;\sigma}\Big|
H_{\mathrm{SO}}
\ket{\underline{q}_\nu;\sigma} &=&  \Big\langle{\underline{k}_\mu}\Big|\bm{K}\ket{\underline{q}_\nu}\cdot \bra{\sigma}\bm{\sigma}\ket{\sigma}  \notag \\
&=& \sigma \Big\langle{\underline{k}_\mu}\Big|{K^z}\ket{\underline{q}_\nu} 
=\sigma\lambda_{k;\mu\nu}^z\delta_{kq}.\hspace*{0.75cm} \label{meth:lz}
\end{eqnarray}
It is straightforward to show that the same result also holds for $N=1, 2$.
Time reversal symmetry requires that
\begin{eqnarray}
\bra{\overline{j}_\mu} H_{\mathrm{SO}} \ket{\overline{i}_\nu}=
(-1)^{i+j-1}
\bra{\overline{-i}_\nu} H_{\mathrm{SO}} \ket{\overline{-j}_\mu}.
\end{eqnarray}
Considering $i=j$ and noting that both are half-odd integers yields  $\lambda_{k;\mu\nu}^z=(\lambda_{k;\nu\mu}^z)^*=-\lambda_{-k;\mu\nu}^z$. Hence  $\lambda_{k;\mu\mu}^z\in\mathbb{R}$, which completes the proof of Eq. (\ref{eq:lz}).

As $\underline\Gamma_{k}$ is a bosonic representation, Eq.  (\ref{meth:lz}) and the orthogonality of the basis functions imply that if ${\cal T}|\underline{k}_\mu\rangle=|\underline{k}_\mu\rangle$ then  $\lambda_{k;\mu\nu}^z=0$  for all $\mu$, $\nu$. Thus $\lambda_{0;\mu\nu}^z=0$ for all $N$ and $\lambda_{N/2;\mu\nu}^z=0$ for even $N$.

Equation (\ref{lxy}) follows similarly on noting that 
\begin{eqnarray}
	\bral{\underline{q}_\nu;\downarrow}
H_{\mathrm{SO}}
\ketl{\underline{k}_\mu;\uparrow} 
=\bral{\underline{q}_\nu}{K^+}\ketl{\underline{k}_\mu}\in\underline\Gamma_{-q}\otimes E_{1}\otimes\underline\Gamma_{k}=A_0
\notag\\
\end{eqnarray}
	 if and only if $k=q-1$.
And that 
\begin{eqnarray}
	\bral{\underline{q}_\nu;\uparrow}
H_{\mathrm{SO}}
\ketl{\underline{k}_\mu;\downarrow} 
=\bral{\underline{q}_\nu}{K^-}\ketl{\underline{k}_\mu}\in\underline\Gamma_{-q}\otimes E_{-1}\otimes\underline\Gamma_{k}=A_0
\notag\\
\end{eqnarray}
	 if and only if $k=q+1$.

\section{Magnetic interactions}\label{magInt}

The parameters for equation (6) are
\begin{eqnarray}
{\cal J}_{\beta\alpha}&=&{\cal J}_{\alpha\beta}, \notag\\
{\cal J}_{xx}&=&{\cal J}_0\left[1+\Lambda^2\cos^2\theta \cos^2\phi\right]+\frac{J}{4},\notag\\
{\cal J}_{yy}&=&{\cal J}_0\left\{1-\Lambda^2\left[ 1+(\cos^2\phi-1)\cos^2\theta \right]\right\}+\frac{J}{4},\notag\\
{\cal J}_{zz}&=&{\cal J}_0\left[1-\Lambda^2\cos^2\theta\right]+\frac{J}{4},\notag\\
{\cal J}_{xy}&=&-{\cal J}_0\Lambda^2\cos^2\theta\cos\phi\sin\phi,\notag\\
{\cal J}_{zx}&=&-{\cal J}_0\Lambda^2\cos\theta\sin\theta \cos\phi,\notag\\
{\cal J}_{yz}&=&{\cal J}_0\Lambda^2\cos\theta\sin\theta \sin\phi,\notag\\
D_y^\pm&=&{\cal J}_0\Lambda\cos\theta\sin\phi,\notag\\
D_z^\pm&=&{\cal J}_0\Lambda\sin\theta \notag\\
\varepsilon_0 &=&-\frac{{\cal J}_0}{4J_c}\left(1+\Lambda^2\right)\left[8\sqrt{t_c^2(1+\Lambda^2)}-3J_c\right]-\frac{J}{16}, \notag
\end{eqnarray}
where $\Lambda=\lambda/t_c$ and
\begin{eqnarray}
{\cal J}_0 &=& \frac{t^2t_c^2J_c}{2(t_c^2+\lambda^2)\left[2(t_c^2+\lambda^2)-J_c\sqrt{(t_c^2+\lambda^2)}\right]}. \notag
\end{eqnarray}
These parameters are  the same regardless of the relative signs of $\lambda$; however,
\begin{eqnarray}
D_x^\pm={\cal J}_0\Lambda(\pm1-\cos\theta\cos\phi)\notag
\end{eqnarray}
is not. In the above $\phi$ is the angle between the local $z$-axes of the molecules and $\theta$ is the angle between the local $y$-axes of the molecules. Relative rotation about the $x$-axis does not change the effective Hamiltonian. The local $x$ and $y$ axes are uniquely determined by the SOC via the phase convention chosen  in equation (3) of the main text.

\end{document}